\documentclass[12pt]{iopart}
\usepackage{iopams}

\newcommand{\CD}{{\cal D}}
\newcommand{\CR}{{\cal R}}
\newcommand{\average}[1]{\left\langle #1 \right\rangle_\CD}

\newcommand{\gaverage}[1]{\left\langle #1 \right\rangle_{\Sigma}}
\newcommand{\initial}[1]{{#1_{\rm \bf i}}}
\newcommand{\inI}{{I}}
\newcommand{\inII}{{II}}

\begin{document}

\title[Globally stationary cosmologies]{On globally static and stationary  cosmologies \\
with or without a cosmological constant \\and the Dark Energy problem}
\author{Thomas Buchert}

\address{Arnold Sommerfeld Center for Theoretical Physics,
Ludwig--Maximilians--Universit\"{a}t, Theresienstra{\ss}e 37,
80333 M\"{u}nchen, Germany \\Email: buchert@theorie.physik.uni-muenchen.de}

\begin{abstract}
In the framework of spatially averaged inhomogeneous cosmologies in classical
General Relativity, effective Einstein equations govern the regional and the global 
dynamics of averaged scalar variables of cosmological models. A particular 
solution may be characterized by a cosmic equation of state. In this paper it is 
pointed out that a globally static averaged dust model is conceivable without 
employing a compensating cosmological constant. Much in the spirit of Einstein's 
original model we discuss consequences for the global, but also for the regional 
properties of this cosmology. We then consider the wider class of globally stationary 
cosmologies that are conceivable in the presented framework. All these models 
are based on exact solutions of the averaged Einstein equations and provide examples 
of cosmologies in an out--of--equilibrium state, which we characterize by an 
information--theoretical measure. It is shown that such cosmologies preserve 
high--magnitude kinematical fluctuations and so tend to maintain their global 
properties. The same is true for a $\Lambda-$driven cosmos in such a state despite 
of exponential expansion. We outline relations to inflationary scenarios, and put  
the Dark Energy problem into perspective. Here, it is argued, on the grounds of the 
discussed cosmologies, that a classical explanation of Dark Energy through 
backreaction effects is  theoretically conceivable, if the matter--dominated Universe 
emerged from a non--perturbative state in the vicinity of the stationary solution.
We also discuss a number of caveats that furnish strong counter arguments in the 
framework of structure formation in a perturbed Friedmannian model.
\end{abstract}



\section{Introduction}

The standard model of cosmology idealizes spatial sections in terms of constant
curvature hypersurfaces, the matter and energy distributions being spatially constant.
Friedmann's and Lema\^\i tre's solutions of Einstein's equations are currently employed to 
describe the dynamics of the Universe as a whole. 
Furthermore, the  assumption is made that the spatially averaged inhomogeneous cosmos is 
described by a member of this family of solutions, a conjecture that can only be proved for 
Newtonian cosmologies \cite{buchertehlers}, \cite{ehlersbuchert}, 
but certainly corresponds to a restricted 
choice in General Relativity.

Friedmann--Lema\^\i tre cosmologies describe
a time--dependent,  locally isotropic and, hence, on a  simply--connected
domain homogeneous and
isotropic expansion or contraction, respectively. Denoting the  scale--factor
by $a(t)$, which can be defined through the volume of an arbitrary comoving 
domain within the space sections, $a:=(V/V(t_i))^{1/3}$, normalized by the volume
at some initial time $t_i$, we obtain from Einstein's equations 
(restricted throughout this paper to a dust continuum) the well--known ``acceleration law'':
\begin{equation}
\label{friedmann1}
\frac{\ddot a}{a} + \frac{4\pi G \varrho_H}{3} -\frac{\Lambda}{3}\;=\;0\;\;,
\end{equation}
with the homogeneous restmass density $\varrho_H$, and the cosmological constant
$\Lambda$ . 
Its first  integral yields a global expansion law, Friedmann's 
differential equation:
\begin{equation}
\label{friedmann2}
\frac{{\dot a}^2 }{a^2 } - \frac{8\pi G\varrho_H}{3}+
\frac{k}{a^2} - \frac{\Lambda}{3}\;=\;0\;\;;\;\;
\varrho_H = \frac{\varrho_H  (t_i)}{a^3}=\frac{M}{V(t_i) a^3}\;\;,
\end{equation}
with the total conserved restmass $M$ enclosed within the arbitrary domain,
and an integration constant $k$ that is related to the spatially constant Ricci 
curvature $R_H$ of the space sections by 
$R_H = 6k/a^2$; $k:=R_H (t_i)/6$\footnote{In Friedmannian cosmology it is
common practice to normalize the curvature parameter to $k:=0,\pm 1$ by a suitable
rescaling of the expansion factor $a(t)$ (then acquiring the dimension of time for
units with $c=1$) . Although we formulated the equations for
a dimensionless scale factor (to be consistent with the general equations to be discussed
below), all numerical estimates in this section and especially in Appendix A 
will be done with the implicit
understanding that $k$ is normalized and $a$ has dimension of time.}

Throughout this paper we shall call a particular solution 
of Friedmann's differential equation (\ref{friedmann2}) a {\it Friedmannian model} or 
a {\it Hubble flow}.
The only possible static Friedmannian model, $a(t)=:a_E$, 
$a_E = const.$, which has motivated 
Einstein \cite{einstein} to introduce the (purely parametric) cosmological term $\Lambda$,
follows from (\ref{friedmann2}) with the assumption that $\dot a = 0$ in some finite time 
interval, and therefore also $\ddot a = 0$, and from Eqs.~(\ref{friedmann1}) and 
(\ref{friedmann2}):
\begin{equation}
\label{einstein1}
4\pi G \varrho_E\;=\;\Lambda\;\;;\;\;\varrho_E := \frac{\varrho_H (t_i)}{a_E^3} 
= const.\;\;;
\end{equation}
\begin{equation}
\label{einstein2}
\frac{k}{a_E^2 } - \frac{8\pi G\varrho_E}{3}  - \frac{\Lambda}{3}
\;=\;0\;\;.
\end{equation}
Combining these two equations we obtain:
\begin{equation}
\label{einstein3}
\frac{k}{a_E^2} \;=\;\Lambda\;\;\;;\;\;\;a_E \;=\;\frac{k}{\sqrt{4\pi G \varrho_E}}\;\;, 
\end{equation}
and since $\varrho_E > 0$, $\Lambda$ has to be {\it positive} and hence also the 
curvature parameter $k$.
Globally, we may introduce the total restmass in the Einstein cosmos 
$M_E= \varrho_E V_E$ with its total volume $V_E := a_E^3 V_i$, and the global scalar
curvature $R_E := 6 k /a_E^2$. Since the curvature is spatially constant and positive, 
the volume can be calculated in terms of spherical space. In space units 
(adopting now the normalization of the curvature parameter to  $k=+1$),
the Riemannian volume of the Einstein cosmos 
is $V_E = 2\pi^2 c^3 a_E^3$, i.e. larger than the
volume of a Euclidean sphere with the same radius $4\pi /3 c^3 a_E^3$. 
However, note that a mass--preserving smoothing of the Einstein radius into a Euclidean geometry
yields a corresponding Euclidean volume $4\pi  /3 c^3 a_E^3 \pi^3$ that is larger than the
total volume of the spherical space of Einstein's cosmos. 
Since $\varrho_E$ is constant, we may rewrite the above equations as follows:
\begin{equation}
\label{einstein4}
4\pi G M_E\;=\;\Lambda V_E\;\;\;;\;\;\;\frac{R_E}{2} + 
8\pi G M_E + \Lambda V_E
\;=\;0\;\;\;.
\end{equation}
$\Lambda V_E$ may be interpreted as the total {\it Dark Energy} in this model.

Einstein's model requires a non--vanishing and positive cosmological constant to
 ``balance'' the total restmass content of the Universe exactly
(including the radiation density and pressure, which we here consider as being negligible
in the matter--dominated era). Note that, by including presssure, 
there also exists a particular static model with $\Lambda =0$ and equation of state
$p_H = -1/3 \varrho_H$
\cite{harrison:book}, p.383). The beauty of a closed spherical space, 
as emphasized by Eddington (\cite{eddington1}, Ch.II),
is accompanied by its definite predictions, e.g. for known restmass density 
we can determine its size as a strong boundary condition for any further studies. 

To illustrate this we calculate in Appendix A the Einstein
radius by extrapolating the values of the cosmological parameters, as fitted to the 
Friedmannian model on the scale of our Hubble volume.
Such estimates are rather naive as will become clear later, and  
we shall come back to this discussion within the more general
setting of an inhomogeneous globally static cosmos. 

Soon after the time when the Einstein static model was suggested, 
the observed redshifts of galaxies 
together with their interpretation as Doppler velocities indicated that the space defined by 
the galaxies in our environment is expanding, which led 
to abandon Einstein's model. This was actually a hasty decision, based on the restricted
view that the {\it global} model by Einstein was asked to 
describe {\it any} patch of the Universe, whatever small it was.
These early discussions were based on the observational situation at the time which,
following Hubble's assessment \cite{hubble}, indicates that the seemingly uniform 
distribution of galaxies may already represent a ``fair sample'' of the Universe. 
As a consequence, these discussions were meant in a global sense: the observed 
-- according to contemporary standards very small -- patch
of the Universe was considered representative for the whole.
Actually, Eddington \cite{eddington2} already pointed out that the observed 
expansion might be a regional property of the Universe rather than a global one; 
he  said {\it ``that it is possible that the recession of the spirals is not the expansion 
theoretically predicted; it might be some local peculiarity masking a much smaller
genuine expansion; but the temptation to identify the observed and the predicted
expansions is very strong''}.

Before the Einstein cosmos was disregarded as a reasonable description of the Universe,
there were many discussions following Einstein's in 1917: Dingle \cite{dingle} and 
Tolman \cite{tolman}, \cite{tolman39} 
have pointed out that the Einstein cosmos will soon develop into a 
highly irregular universe. This discussion is still referred to as the ``instability of the 
Einstein cosmos''. Eddington \cite{eddington2} argued that 
the Universe may evolve starting out from the Einstein cosmos, however, due to its 
instability, will start expanding or contracting, respectively.
As an alternative, Lema\^\i tre \cite{lemaitre33a}, \cite{lemaitre33b} advocated the
singular ``Big Bang solutions'' which expand until the 
matter density has dropped below the cosmological term which henceforth dominates, 
resulting in an accelerating phase thereafter (see also \cite{peebles} \S 3C for 
a review of these discussions). The nowadays favoured {\it concordance model} 
({\it cf.} Appendix A), 
featuring a positive cosmological constant,  describes such an evolution.
We shall see that the ideas advanced at that time
apply to the picture developed below, but their interpretation will be very different. 
In contrast to the historical flaw of realizing the instability of the Einstein cosmos, 
we nowadays view such instabilities as the origin of large--scale structure. 
All homogeneous world models are unstable including the Einstein cosmos.
As we shall discuss in detail below, the {\it global} instability argument in the above
form (i.e., within the class of homogeneous--isotropic models)
does not apply, when the cosmology acquires the status of describing the 
average dynamics on the largest scales. 

Here, a disclaimer is in order: to advance a globally static, but regionally fluctuating 
cosmos as a viable model that could explain current observational results is premature. 
Instead, we revisit the ideas which led to the Einstein cosmos (the introduction of the
cosmological constant) in light of a new framework and on the grounds of an ongoing
discussion of the possibility that {\it Dark Energy} may be explained through 
``backreaction effects'' of structure formation. Globally expanding,  stationary 
cosmologies are also conceivable in this framework. It was pointed out in 
\cite{buchert:darkenergy} that  the question whether the magnitude of 
backreaction effects is sufficient
to explain acceleration of the observable Universe, and the question of 
whether a globally static or stationary cosmos bears physical justifications 
beyond a mere mathematical possibility are synonyms. 

A thorough physical investigation of fluctuation--supported static and stationary 
cosmologies must be 
based on more general matter models (as a next step perfect fluids and scalar fields) 
that allow to study
dynamical scenarios of inflation and their exit details including the effect of 
radiation pressure. The corresponding effective equations are 
given in \cite{buchert:grgfluid}. In this respect the present investigation based on the
effective equations  for a dust continuum \cite{buchert:grgdust} provides a 
useful showcase for the presentation of the basic ideas. The present investigation of
particular exact solutions offers more insight into the general formalism
of averaged inhomogeneous cosmologies, and at the same time proposes new families of
cosmologies that enjoy significantly more freedom than a rigid Friedmannian cosmology. 
Furthermore, by applying these ideas to inflationary scenarios, we can understand the
relevance of  matter and curvature fluctuations for the description of the Early Universe, in
particular the importance of the role played by a non--vanishing averaged scalar 
curvature.

\section{Effective equations for inhomogeneous universe models}

\subsection{Averaged equations}

For the sake of transparency we shall restrict all considerations in this paper to 
an irrotational dust continuum and recall a set of effective equations provided in 
\cite{buchert:grgdust}. The ideas presented can be carried over to studies of 
inhomogeneous cosmologies covering the Early Universe and radiation--dominated epochs
with the help of the more general effective equations developed in \cite{buchert:grgfluid}.

Given a foliation of  spacetime into flow--orthogonal hypersurfaces (which is possible for
irrotational dust) with the 3--metric $g_{ij}$ in the line--element 
$ds^2 = -dt^2 + g_{ij}\,dX^i dX^j$,
spatial averaging of any scalar field $\Psi$ is a covariant operation and is defined 
by the simple averager:
\begin{equation}
\label{eq:average-GR}
\average{\Psi (t, X^i)}: = 
\frac{1}{V_\CD}\int_\CD J d^3 X \;\;\Psi (t, X^i) \;\;,
\end{equation}
with  $J:=\sqrt{\det(g_{ij})}$; $g_{ij}$ is  the metric  of the
spatial  hypersurfaces, and  $X^i$ are  coordinates that  are constant
along  flow lines, which are here spacetime geodesics. 
Following \cite{buchert:grgdust} we define an 
{\it effective scale factor} by the volume of a simply--connected domain 
$\CD$ in a t--hypersurface, normalized by the  volume of the initial domain $\initial\CD$,
\begin{equation}
a_\CD := \left(\frac{V_\CD}{V_{\initial\CD}}\right)^{1/3}\;\;.
\end{equation}
We recall the fact that, for a restmass preserving domain $\CD$, 
volume averaging of a scalar function $\Psi$
does not commute with its time--evolution:
\begin{equation}
\label{commutativity}
\average{\partial_t \Psi} - \partial_t 
{\average{\Psi}}\;=\;\average{\Psi}\average{\theta} -
\average{\Psi\theta}\;\;,
\end{equation}
where $\theta$ denotes the rate of expansion.
Setting $\Psi \equiv \varrho$ we obtain a regional continuity equation reflecting 
the conservation of the total restmass $M_\CD$ within $\CD$:
\begin{equation}
\label{massconservation}
\partial_t M_\CD \;=\;0\;\;\;\Leftrightarrow\;\;\; \partial_t \average{\varrho} + \average{\theta}
\average{\varrho}\;=\;0\;\;.
\end{equation}
Setting $\Psi \equiv \theta$
we can derive an effective equation for the spatially averaged expansion of the model,
\begin{equation}
\label{effectivehubble}
\average{\theta} \;=\;\frac{{\dot V}_\CD}{V_\CD}\;=\;3\frac{{\dot a}_\CD }{a_\CD }\;=\;: 3 H_\CD \;\;, 
\end{equation}
where we defined an {\it effective Hubble functional} on $\CD$
(an overdot denotes partial time--derivative): 
inserting {\it Raychaudhuri's evolution equation}, 
${\dot\theta} = \Lambda - 4\pi G \varrho -\frac{1}{3}\theta^2 - 2 \sigma^2$
(with the rate of shear $\sigma^2 = \frac{1}{2}\sigma_{ij}\sigma^{ij}$), into 
(\ref{commutativity}) and using the effective scale--factor $a_\CD$ we obtain:
\begin{equation}
\label{averageraychaudhuri}
3\frac{{\ddot a}_\CD}{a_\CD} + 4\pi G 
\frac{M_\CD}{\initial{V} a_\CD^3} - \Lambda = {\cal Q}_\CD\;\;.
\end{equation}
The first integral of the above equation is directly given by averaging 
the {\it Hamiltonian constraint}:
\begin{equation}
\label{averagehamilton}
\left( \frac{{\dot a}_\CD}{a_\CD}\right)^2 - \frac{8\pi G}{3}
\frac{M_\CD}{\initial{V}a_\CD^3} + \frac{\average{\CR}}{6} 
- \frac{\Lambda}{3} = -\frac{{\cal Q}_\CD}{6} \;\;,
\end{equation}
where   the  total restmass   $M_\CD$,   the  averaged   spatial  Ricci   scalar
$\average{\CR}$   and   the  {\it kinematical backreaction  term}  ${\cal Q}_\CD$   are
domain--depen\-dent and, except the mass, time--depen\-dent functions.
The backreaction source term is given by
\begin{equation}
\label{eq:Q-GR} 
{\cal Q}_\CD : = 2 \average{\inII} - \frac{2}{3}\average{\inI}^2 =
\frac{2}{3}\average{\left(\theta - \average{\theta}\right)^2 } - 
2\average{\sigma^2}\;\; ;
\end{equation}
here,  $\inI = \Theta^i_{\;i}$  and $\inII = \frac{1}{2}[\,(\Theta^i_{\;i})^2 - 
\Theta^i_{\;j}\Theta^j_{\;i}\,]$  
denote  the  principal scalar invariants  of the  expansion
tensor, defined  as minus the extrinsic
curvature  tensor $K_{ij}:=\Theta_{ij}$. In the second equality above it was split 
into kinematical invariants through the decomposition  
$\Theta_{ij} = \frac{1}{3}g_{ij}\theta + \sigma_{ij}$, with the rate of expansion 
$\theta =\Theta^i_{\;i}$, and the shear tensor $\sigma_{ij}$.
(Note that vorticity is absent in
the present model; we adopt the summation convention.)

The time--derivative of the averaged Hamiltonian constraint (\ref{averagehamilton})
agrees with the averaged Raychaudhuri equation (\ref{averageraychaudhuri}) by virtue of the
following {\it integrability  condition}:
\begin{equation}
\partial_t {\cal Q}_\CD + 6 \frac{{\dot a}_\CD}{a_\CD} {\cal Q}_\CD +  
\partial_t \average{\CR}
+ 2 \frac{{\dot a}_\CD}{a_\CD} \average{\CR} = 0 \;\;,
\end{equation}
which we may write in the more compact form:
\begin{equation}
\label{integrability}
\frac{1}{a_\CD^6}\partial_t \left(\,{\cal Q}_\CD a_\CD^6 \,\right) 
\;+\; \frac{1}{a_\CD^{2}} \;\partial_t \left(\,\average{\CR}a_\CD^2 \,
\right)\,=0\;.
\end{equation}
Formally integrating this condition yields: 
\begin{equation}
\label{integrabilityintegral}
\frac{k_\CD}{a_\CD^2} - \frac{1}{3 a_\CD^2} \int_{t_i}^t \,dt' \;
{\cal Q}_\CD\; \frac{d}{dt'} a^2_\CD(t')
= \frac{1}{6}\left(\,\average{\CR} + {\cal Q}_\CD\,\right) \;\;,
\end{equation}
i.e., besides the total material mass $M_\CD$ we have a further integral of motion given by
the domain--dependent integration constant 
$k_\CD$ that we may also write as follows:
\begin{equation}
\label{yamabe}
6\,k_\CD = Y_\CD+ 2 \int_{t_i}^t \,dt' \;
{\cal Q}_\CD\; \frac{d}{dt'} a^2_\CD(t') -{\cal Q}_\CD a_\CD^2\;\;.
\end{equation}
In the above equation we introduced the functional $ Y_\CD$, which is a special case of the
{\it Yamabe functional}, here $Y_\CD := \average{\CR}V_\CD^{2/3}$ in three dimensions
(see, e.g., the case $n=3; \phi = \psi = 1$ in \cite{aubin}, p.150), 
which itself is an integral of motion
for vanishing ${\cal Q}_\CD$\footnote{The vanishing of ${\cal Q}_\CD$ on every scale is necessary  
and sufficient for $a_\CD$ to be a (global) solution of Friedmann's differential equation.
Therefore, a non--vanishing ${\cal Q}_\CD$ plays a key--role and justifies the name
``kinematical backreaction''. $Y_\CD$ is also preserved in the special case where 
${\cal Q}_\CD \propto a_{\cal D}^{-6}$, which is briefly discussed at the end of this
section.}.
Eq.~(\ref{integrability}), having no Newtonian analogue,
shows  that  the averaged intrinsic curvature  and  the averaged extrinsic curvature 
(encoded in the backreaction  term) are dynamically coupled. 
Stating this genuinly relativistic property, we also note the surprising fact that,
inserting (\ref{integrabilityintegral}) into (\ref{averagehamilton}) results in an equation that
is formally equivalent to its Newtonian counterpart \cite{buchertehlers}:
\begin{equation}
\label{averagefriedmann}
\frac{\dot{a}_\CD^2 + k_\CD}{a_\CD^2 } - \frac{8\pi G \average{\varrho}}{3}
- \frac{\Lambda}{3} = \frac{1}{3 a_\CD^2} \int_{\initial{t}}^t \rmd t'\ {\cal Q}_\CD
\frac{\rmd }{\rmd t'} a^2_\CD(t')\;\;.
\end{equation}
The effective scale--factor obeys the same equation as in Newtonian theory similar 
to the situation known for the homogeneous--isotropic case. 
Note that these effective equations also cover anisotropic inhomogeneous 
cosmologies (\cite{buchertehlers}, Appendix B).

\subsection{The cosmic quartet}

For the purpose of comparing the model variables with observations  it is comfortable to 
introduce dimensionless average characteristics as follows (contrary to \cite{buchert:grgdust}
we use the notation $\Omega_{\cal R}^{\CD}$ for the curvature functional):
\begin{equation}
\label{omega}
\fl\qquad
\Omega_m^{\CD} : = \frac{8\pi G M_{\CD}}{3 V_{\initial{\CD}}a_{\CD}^3 
H_{\CD}^2 } \;\;;\;\;
\Omega_{\Lambda}^{\CD} := \frac{\Lambda}{3 H_{\CD}^2 }\;\;;\;\;
\Omega_{\cal R}^{\CD} := - \frac{\average{\cal R}}{6 H_{\CD}^2 }\;\;;\;\;
\Omega_{\cal Q}^{\CD} := - \frac{{\cal Q}_{\CD}}{6 H_{\CD}^2 } \;\;,
\end{equation}
where we have employed the  effective Hubble--functional $H_\CD$ (\ref{effectivehubble})
that reduces to Hubble's function in the homogeneous--isotropic case.
With these definitions the Hamiltonian constraint is written in the  iconized form of
a  {\it cosmic quartet}:
\begin{equation}
\label{hamiltonomega}
\Omega_m^{\CD}\;+\;\Omega_{\Lambda}^{\CD}\;+\;\Omega_{\cal R}^{\CD}\;+\;
\Omega_{\cal Q}^{\CD}\;=\;1\;\;.
\end{equation}
These functionals, being scale--dependent,
are dynamically related in a complex way\footnote{Here, the
detailed investigation of these functionals in the framework of Newtonian cosmology 
in \cite{bks} is useful to understand this remark.}, 
unlike the situation in a Friedmannian model that features a global
{\it cosmic triangle} \cite{bahcall}
where the cosmological parameters interact trivially, e.g. for 
an Einstein--de Sitter cosmology (vanishing curvature parameter and vanishing $\Lambda$)
the curvature parameter remains zero throughout the entire evolution. 

As in Friedmannian cosmology we may also define other functionals, as for example
an {\it effective deceleration parameter},
\begin{equation}
\label{deceleration}
q_{\rm eff}^\CD := -\frac{{\ddot a}_\CD}{a_\CD}\frac{1}{H_\CD^2} = \frac{1}{2}
\Omega_m^{\CD} - \Omega_{\Lambda}^{\CD} + 2 \Omega_{\cal Q}^{\CD}\;\;,
\end{equation}
as well as corresponding effective parameters for the third derivative of $a_\CD$ 
like the recently introduced {\it state finders} \cite{alametal} (see also \cite{evans} and 
references therein).

We also note the useful general evolution equation (a slightly different version is given in 
\cite{buchert:grgdust}, Appendix B):
\begin{equation}
\label{evolutionomega}
\fl
\frac{1}{a_\CD^6}\partial_t \left(\, \Omega_{\cal Q}^{\CD}a_\CD^6\,\right)  
\;+\; \frac{1}{a_\CD^{2}} \;\partial_t \left(\,\Omega_{\cal R}^{\CD}a_\CD^2 \,\right)\,-
\,3H_\CD \left(\, \Omega_{\cal Q}^{\CD}+ \Omega_{\cal R}^{\CD}  \,\right) \,\left(\,
\Omega_m^{\CD} + \frac{2}{3} \Omega_{\cal R}^{\CD} + 
2 \Omega_{\cal Q}^{\CD}\,\right)\,=\,0 \;.
\end{equation}
 
A clarifying remark concerning the definition of the ``curvature parameter'' is in order.
In the corresponding Newtonian problem the following curvature and kinematical
backreaction functionals have been introduced \cite{bks}:
\begin{equation}
\label{omeganewton}
\Omega_{k}^{\CD} := - \frac{k_\CD}{a_\CD^2 H_{\CD}^2 }\;\;;\;\;
\Omega_{{\cal Q}N}^{\CD} := \frac{1}{3 a_\CD^2 H_\CD^2}
\int_{\initial{t}}^t \rmd t'\ {\cal Q}_\CD\frac{\rmd }{\rmd t'} a^2_\CD(t')\;\;.
\end{equation}
In view of Eq.~(\ref{averagefriedmann}) we may use the same functionals in place of those
introduced above also in General Relativity. However, the physical averaged curvature is
not associated with $\Omega_k^\CD$. The functionals (\ref{omeganewton}) are related to 
the previously introduced ones by: $\;\Omega_{k}^{\CD} +\Omega_{{\cal Q}N}^{\CD}\;=\; 
\Omega_{\cal R}^{\CD} + \Omega_{\cal Q}^{\CD}$.

\subsection{The cosmic equation of state}

The above equations can formally be recast into standard zero--curvature Friedmann equations 
with new effective sources \cite{buchert:grgfluid}\footnote{Note that in this representation
of the effective equations ${p}_{\rm eff}$ just denotes
a formal ``pressure'':  in the perfect fluid case with an inhomogeneous pressure function
the foliation has to be differently chosen (the in general inhomogeneous lapse function is set equal
to $1$ here),  and there is a further averaged pressure gradient
term \cite{buchert:grgfluid}.}:
\begin{equation}
\label{equationofstate}
\fl
\varrho^{\CD}_{\rm eff} = \average{\varrho}-\frac{1}{16\pi G}{\cal Q}_\CD - 
\frac{1}{16\pi G}\average{\CR}
\;\;\;;\;\;\;{p}^{\CD}_{\rm eff} =  -\frac{1}{16\pi G}{\cal Q}_\CD + \frac{1}{48\pi G}\average{\CR}\;\;.
\end{equation}
\begin{equation}
\label{effectivefriedmann}
\fl
3\frac{{\ddot a}_\CD}{a_\CD} =
\Lambda - 4\pi G (\varrho^{\CD}_{\rm eff}
+3{p}^{\CD}_{\rm eff})\;\;;\;\;
3H_\CD^2 =\Lambda + 8\pi G 
\varrho^{\CD}_{\rm eff}\;\;;\;\;
{\dot\varrho}^{\CD}_{\rm eff} + 
3H_\CD \left(\varrho^{\CD}_{\rm eff}
+{p}^{\CD}_{\rm eff} \right)=0\;.
\end{equation}
Eqs.~(\ref{effectivefriedmann}) correspond to the equations (\ref{averageraychaudhuri}),
(\ref{averagehamilton}) and (\ref{integrability}), respectively. 
In these equations we have translated all what has been said before into a Friedmannian setting
and a specific form of the fluctuating sources.
Note that the kinematical backreaction term ${\cal Q}_\CD$ itself 
obeys a {\it stiff} equation of state mimicking a dilatonic field in the fluid analogy
(for further implications see \cite{buchert:grgfluid} and Subsect.~\ref{subsect:inflation}).

Given an equation of state of the form $p^\CD_{\rm eff} = 
\beta\; (\varrho^\CD_{\rm eff}, a_{\cal D})$
that relates the effective sources (\ref{equationofstate}) with a possible explicit
dependence on the effective scale factor, 
the effective Friedmann equations (\ref{effectivefriedmann})
can be solved (one of the equations (\ref{effectivefriedmann}) is redundant). 
Therefore, any question posed that is related to the evolution of scalar characteristics
of inhomogeneous universe models may be ``reduced'' to finding the {\it cosmic state}
on a given spatial scale. 
(Note, however, that an equation of state must not exist in general.)
Although
formally  similar to the situation in Friedmannian cosmology, here the equation of state
is {\it dynamical} and depends on details of the evolution of inhomogeneities.
In general it describes non--equilibrium states.

An example may illustrate the {\it cosmic equation of state}: we look at a particular 
exact solution of the averaged Einstein equations that was given in (\cite{buchert:grgdust}
Appendix B). From the integrability condition (\ref{integrability}) we directly infer that 
the pair of solutions,
\begin{equation}
\label{decouplingsolution}
{\cal Q}_{\cal D} \;=\; \frac{{\cal Q}_{\cal D}(t_i)}{a_{\cal D}^{6}}\;\;\;;\;\;\;
\average{\CR} \;=\;  \frac{\average{\CR}(t_i)}{a_{\cal D}^{2}}\;\;,
\end{equation}
provides a special solution for which the averaged scalar curvature and the kinematical
backreaction term decouple and evolve independently. The regional cosmic equation of
state corresponding to this solution can be derived from (\ref{equationofstate}):
\begin{equation}
\label{equationofstatedecoupling}
\frac{p^\CD_{\rm eff}}{\varrho^\CD_{\rm eff}}=:w^\CD_{\rm eff} =
\frac{1-\frac{1}{3}\gamma_1 a_{\cal D}^4}{1+\gamma_1 a_{\cal D}^4 - \gamma_2 
a_{\cal D}^3}\;\;,
\end{equation}
with $\gamma_1 := \Omega^\CD_{\cal R}(t_i)/\Omega^\CD_{\cal Q}(t_i)$ and
$\gamma_2 :=  \Omega^\CD_{m}(t_i)/\Omega^\CD_{\cal Q}(t_i)$.
Asymptotically, for an expanding domain ($|a_{\cal D}|$ large), $w^\CD_{\rm eff}$ tends to
$-\frac{1}{3}$, a property that is also shared by a non--flat Friedmannian domain for which
we obtain
$w^{\rm Friedmann} = -\frac{1}{3}/(1+ \gamma_3 /a_{\cal D})$ with
$\gamma_3 :=\Omega^\CD_{m}(t_i)/\Omega^\CD_{\cal R}(t_i)$.  
These Friedmannian models
are subcases of the above inhomogeneous solution ($\Omega^\CD_{\cal R}(t_i) \ne 0; 
\Omega^\CD_{\cal Q}(t_i) =0$); $w^{\rm Friedmann}$ follows from 
(\ref{equationofstatedecoupling}) by first multiplying with 
$\Omega^\CD_{\cal Q}(t_i) \ne 0$ (it is understood that the scale--factor and all 
other quantities then no longer depend on $\cal D$).

We end this section with a note on the definition of the effective sources 
in Eq.~(\ref{equationofstate}):
there is some ambiguity in defining them. Firstly, it may
sometimes be useful to incorporate $\Lambda$ into the effective 
sources by defining $\varrho^\CD_{\rm eff\Lambda}: = \varrho^\CD_{\rm eff} - \Lambda/8\pi G$
and $p^\CD_{\rm eff\Lambda} := p^\CD_{\rm eff} + \Lambda/8\pi G$.
Secondly, we might add the ``constant curvature term'' $3 k_\CD / a_\CD^2$
to the left--hand--side of the second equation in (\ref{effectivefriedmann}); if we wish to do
so, then the effective sources can be represented solely through the kinematical
backreaction term ${\cal Q}_\CD$ and its time--integral. For this we have to exploit
the ``Newtonian form'', Eq.~(\ref{averagefriedmann}), and would have to define the 
effective sources as follows:
\begin{equation}
\label{equationofstate-k}
\fl
\hat{\varrho}^{\CD}_{\rm eff}: = \average{\varrho}+\frac{X_\CD}{16\pi G} 
\;\;;\;\;{\hat p}^{\CD}_{\rm eff}: =  -\frac{{\cal Q}_\CD}{12\pi G} - \frac{X_\CD}{48\pi G}
\;\;;\;\;  X_\CD := \frac{2}{ a_\CD^2} \int_{\initial{t}}^t \rmd t'\ {\cal Q}_\CD
\frac{\rmd }{\rmd t'} a^2_\CD(t')  \;.
\end{equation}
The integrated form of the integrability condition, Eq.~(\ref{integrabilityintegral}), then
allows to express $X_\CD$ again through the averaged scalar curvature, 
$X_\CD = 6k_\CD - {\cal Q}_\CD - \average{\cal R}$, and we obtain the sources 
corresponding to (\ref{equationofstate}), however, with a curvature source that 
captures the deviations from a constant curvature model:
\begin{equation}
\label{equationofstate-kk}
\fl
\hat{\varrho}^{\CD}_{\rm eff} = \average{\varrho}-\frac{{\cal Q}_\CD}{16\pi G} - 
\frac{\left[\average{\CR}-6 k_\CD / a_\CD^2 \right]}{16\pi G}
\;\;;\;\;{\hat p}^{\CD}_{\rm eff} =  -\frac{{\cal Q}_\CD}{16\pi G} + 
\frac{\left[\average{\CR}-6 k_\CD / a_\CD^2 \right]}{48\pi G}\;.
\end{equation}

\vspace{10pt}

In the following section we present new families of exact solutions of the effective 
Einstein equations. While models can be obtained on any chosen simply--connected
spatial domain $\CD$, most of the following solutions are {\it global} and the   
spatial domain $\CD$ is extended to the whole Riemannian manifold $\Sigma$, 
which we assume to be compact. (Note that {\it regional solutions} such as the example
presented in Eq.~(\ref{decouplingsolution}) ``appear'' to only depend on the matter
distribution inside $\CD$, but this is not the case, since the initial data are to be constructed
non--locally from the whole distribution in $\Sigma$; global solutions and their parameters
therefore have a more robust status. 

\section{Globally stationary effective universe models}

\subsection{Globally static cosmos without a cosmological constant}

On the global scale we first require the effective scale--factor $a_{\Sigma}$ to be 
constant on some time--interval, hence ${\dot a}_{\Sigma} = {\ddot a}_{\Sigma}=0$ and
Eqs.~(\ref{averageraychaudhuri}) and (\ref{averagehamilton}) 
may be written in the form:
\begin{equation}
\label{static1}
{\cal Q}_{\Sigma}\;=\;4\pi G \frac{M_{\Sigma}}{\initial{V}a_{\Sigma}^3} - \Lambda\;\;;
\end{equation}
\begin{equation}
\label{static2}
\gaverage{\CR}\;=\; 12\pi G \frac{M_{\Sigma}}{\initial{V}a_{\Sigma}^3}+3\Lambda\;\;,
\end{equation}
with the global {\it kinematical backreaction} ${\cal Q}_{\Sigma}$, the globally averaged 3--Ricci 
curvature $\gaverage{\CR}$, and the total restmass  $M_{\Sigma}$ contained in
$\Sigma$.

Let us now consider the case of a vanishing cosmological constant: $\Lambda = 0$.
The averaged scalar curvature  is, for a non--empty Universe, 
always {\it positive}, and 
the balance condition (\ref{static1}) replaces (\ref{einstein1}), while 
the condition (\ref{static2}) replaces (\ref{einstein2}). 
Obviously, backreaction (\ref{static1}) and averaged scalar curvature (\ref{static2}) 
trivially satisfy the integrability condition (\ref{integrability}).
Thus, in view of (\ref{equationofstate}), 
the globally static inhomogeneous cosmos without a cosmological constant 
is characterized by the cosmic equation of state:
\begin{equation}
\label{cosmicstate1}
\gaverage{\CR}\;=\;3{\cal Q}_{\Sigma}\;=\;const.\;\;\Rightarrow\;\;
{p}^{\Sigma}_{\rm eff}\;=\; \varrho^{\Sigma}_{\rm eff}\;=\;0\;\;.
\end{equation}

\subsection{Interlude: kinematical backreaction as  a cosmological constant}

We first note that, apparently, in Eq.~(\ref{averageraychaudhuri})
${\cal Q}_\CD$ plays the role of a positive cosmological constant; however,  in 
Eq.~(\ref{averagehamilton}) ${\cal Q}_\CD$ has opposite sign! That the 
kinematical backreaction term may take the role of the cosmological 
constant has been suggested in \cite{buchert:jgrg} and
discussed in connection with the backreaction problem in \cite{bks} as well as in the
recent discussion on Dark Energy and backreaction \cite{kolbetal,rasanen:constraints}. 
The above remark shows that caution is in order with a direct identification.

Of course, we may {\it force} the kinematical backreaction term to take exactly the role
of the cosmological constant: neglecting $\Lambda$ in the general equation 
(\ref{averageraychaudhuri}), ${\cal Q}_\CD$ may be regarded as a (in this equation 
possibly time--dependent) cosmological term. Then, from Eq.~(\ref{averagehamilton})
a constraint equation on  ${\cal Q}_\CD$ follows, if we identify all sources that 
imply a deviation from Friedmann's equation with the cosmological term; the correct
requirement can be inferred from Eq.~(\ref{averagefriedmann}) and reads:
\begin{equation}
\label{lambdacondition}
 \frac{2}{ a_\CD^2 }
\int_{\initial{t}}^t \rmd t'\ {\cal Q}_\CD\frac{\rmd }{\rmd t'} a^2_\CD(t')
\;\equiv\;{\cal Q}_\CD \;\;,
\end{equation}
which implies ${\cal Q}_\CD = {\cal Q}_{\cal D}(t_i) = const.$ as the only possible 
solution. Physically, it appears contrived to freeze fluctuations to a constant value
in an evolving model.
However, it is noteworthy that ${\cal Q}_\CD$ indeed is required to be constant, if 
we force it to behave like $\Lambda$.
Eq.~(\ref{integrabilityintegral}) then implies for the averaged curvature:
\begin{equation}
\label{lambdacurvature}
\average{\cal R} \;=\;\frac{6 k_\CD}{a_\CD^2} \;-\;3 {\cal Q}_\CD (t_i)\;\;.
\end{equation}
The solution in the case $k_\CD =0$ has been noticed in \cite{kolbetal} too; however,
freezing also the averaged curvature in an evolving model appears to be even more
contrived. A different interpretation in the framework of an effective scalar field 
may imply a more meaningful interpretation of this solution (see 
Subsect.~\ref{subsect:inflation}). We finally write the 
cosmic equation of state for this special (regional) solution:
\begin{equation}
\label{lambdacosmic}
\frac{p^\CD_{\rm eff}}{\varrho^\CD_{\rm eff}} \;=\;
-\;\frac{{\cal Q}_{\cal D}(t_i) - k_\CD / a_\CD^2}{8\pi G \average{\varrho}(t_i)/a^3_\CD +   
{\cal Q}_{\cal D}(t_i) - 3 k_\CD / a_\CD^2} \;,
\end{equation}
which, for large $|a_\CD |$, approaches the equation of state 
$p^\CD_{\rm eff} = -\varrho^\CD_{\rm eff}$.

\subsection{Local instability versus global stability}

The problem of instability of the Einstein cosmos, as outlined in the introduction, 
must be thought of in two ways: i) a static homogeneous cosmos is unstable 
within the class of Friedmann--Lema\^\i tre cosmologies, 
since for a small change in density the balance
between the density and the cosmological constant in Eq.~(\ref{einstein1}) is destroyed, 
leading to acceleration if $4\pi G \varrho_H (t_i) > \Lambda$, and to decceleration otherwise.
Suppose we take a slightly smaller density, then Eq.~(\ref{einstein2})  implies 
that $H (t) > 0$, and the model starts to expand;
$\Lambda$ (being constant in time) cannot dynamically compensate for this expansion,
which itself decreases the density further, and so gives rise to the instability;
ii) a  homogeneous cosmos is unstable to inhomogeneous density perturbations.
Such perturbations are amplified as a consequence of the attractive nature of the 
gravitational field tending to increase overdensities and to decrease underdensities. 
This is the content of {\it local gravitational instability}: inhomogeneities are amplified.
The latter instability also applies to the other Friedmann--Lema\^\i tre cosmologies.
Both types of scalar instabilities (and also vectorial and tensorial perturbations)
have been recently clarified and detailed by 
Barrow et al. \cite{barrowetal:static} for the Einstein static universe model containing
a perfect fluid,  generalizing earlier work by Harrison \cite{harrison} and Gibbons
\cite{gibbons87}, \cite{gibbons88}. Further insight was added by 
Losic and Unruh \cite{unruh:static} 
who investigated the stability analysis to second order in 
scalar and metric fluctuations. 
A related analysis concerning the dynamical phase space
as well as  attractor or
repellor properties of homogeneous solutions may be found in \cite{ehlersrindler},
\cite{sota:RG}.

Now, let us look at the same type of perturbations in the framework of the globally
static, but inhomogeneous cosmos. 
Altering the density source would equally disturb the balance
by virtue of Eq.~(\ref{static1}), but it would not necessarily destroy it: the reason is
that the kinematical backreaction term ${\cal Q}_{\Sigma}$ 
indeed reacts back on this perturbation (justifying its name). 
For, unlike $\Lambda$, ${\cal Q}_{\Sigma}$ 
can acquire a time--dependency and is then dynamically
coupled to the averaged scalar curvature $\gaverage{\CR}$ (in general 
through the integrability condition (\ref{integrability}) and in particular, 
for a stationary cosmos, through Eq.~(\ref{coupling}) below):
as soon as the density is perturbed, taking for example 
a slightly lower average density as above, the (positive)
curvature starts to evolve such that 
$\partial_t \,(\, \gaverage{\CR} V_\CD^{2/3} \,)<0$; 
in turn, 
$\partial_t \,(\,{\cal Q}_{\Sigma}V_\CD^2 \,) > 0$, 
and, thus, fluctuations tend to decrease less rapid: as we shall see in an exact solution
below, the coupling to the averaged scalar curvature can be strong so that 
${\cal Q}_{\Sigma}$ decreases in proportion to the inverse volume similar to the
density, not (as expected for fluctuations) in proportion to the square of the inverse
volume.
It does not matter if we consider homogeneous or inhomogeneous perturbations
(as far as scalar perturbations are concerned), since
the effective equations (\ref{averageraychaudhuri}) and (\ref{averagehamilton}) 
govern both and are not narrowed to the class of homogeneous--isotropic
solutions as in the case of the standard Einstein cosmos.
In other words: a perturbed effective cosmology stays within the same class of cosmologies
governed by the effective equations, since the latter are general.

An interesting future task will be to analyze in detail -- along the lines of a stability
analysis of the homogeneous cosmos with symmetry \cite{barrowetal:static}, 
\cite{unruh:static} -- whether 
kinematical backreaction tends to stabilize perturbations of the dynamical 
balance between the globally averaged density source $4\pi G \gaverage{\varrho}$
and the global kinematical backreaction term  ${\cal Q}_{\Sigma}$ 
due to its coupling to the averaged 3--Ricci curvature $\gaverage{\cal R}$.
In any case, we do not expect that a generic cosmology would dynamically approach
this state for any initial setting. In particular, a perturbed Friedmannian state
may not approach a globally static state, rather the question is whether 
perturbations of an {\it already established} balance  would 
destabilize this state. This calls for an investigation of perturbation theory of the global
out--of--equilibrium state, rather than of a Friedmannian state.

\subsection{Globally stationary effective cosmologies}

Suppose that the Universe indeed is hovering around 
a non--accelerating state on the largest scales. Still, the effective static cosmos 
discussed above may appear as a quite rigid model. We may think that a more
natural condition would be {\it stationarity}. Indeed, at first sight the  
balance condition, if attained, does not necessarily imply that the model is static. 
We may look for a wider class of models that balances the fluctuations and the 
averaged sources by introducing 
{\it globally stationary effective cosmologies}: the vanishing of the second time--derivative
of the scale--factor would only imply ${\dot a}_{\Sigma} = const.=:{\cal C}$, i.e.,
$a_{\Sigma} = a_S + {\cal C}(t-t_i)$, where the integration constant $a_S$
is generically non--zero, e.g. the model may emerge \cite{ellis:emergent1}, \cite{ellis:emergent2}
from a globally static cosmos, $a_S :=1$, or from a  
`Big--Bang', if $a_S$ is set to zero. (In Friedmannian cosmology such an expansion law
would correspond to a ``curvature--dominated'' model, since ${\cal C}^2 + k \approx 0$
for negligible sources, hence resulting in a Hubble expansion that is determined by 
a constant negative curvature.)

On the global scale and for a stationary 
cosmos, Eqs.~(\ref{averageraychaudhuri}) and (\ref{averagehamilton}) read:
\begin{equation}
\label{stationary1}
{\cal Q}_{\Sigma}\;=\;4\pi G \frac{M_{\Sigma}}{\initial{V}a_{\Sigma}^3} - \Lambda\;\;;
\end{equation}
\begin{equation}
\label{stationary2}
\gaverage{\CR} \;=\;12\pi G \frac{M_{\Sigma}}{\initial{V}a_{\Sigma}^3}  + 3\Lambda- 
6H_{\Sigma}^2  \;\;;\;\;H_{\Sigma}=\frac{\cal C}{a_{\Sigma}}\;\;.
\end{equation}
By inserting (\ref{stationary1}) into (\ref{stationary2}), we can evaluate the 
constant $\cal C$ by looking at the resulting equation at initial time; for the normalization
$a_{\Sigma}(t_i) =1$ we get:
\begin{equation}
\label{constantC}
6{\cal C}^2 = 6\Lambda + 3{\cal Q}_{\Sigma}(t_i) - \gaverage{\CR}(t_i)\;\;.
\end{equation}
We are now going to discuss the stationary 
cosmologies more explicitly by deriving an exact solution to the effective Einstein equations.
(One can easily show from (\ref{friedmann2}) that a Friedmannian cosmology does not
allow for a stationary cosmos, since ${\cal C}^2 + k = \Lambda a^2$ only allows
for $a=const$.)

First, note that the time--derivatives of the above equations 
deliver a dynamical coupling relation between  
${\cal Q}_{\Sigma}$ and $\gaverage{\CR}$ as a special case of the integrability condition
(\ref{integrability}):
\begin{equation}
\label{coupling}
-\partial_t {\cal Q}_{\Sigma} + \frac{1}{3} \partial_t  \gaverage{\CR}\;=\;
\frac{4{\cal C}^3}{a_{\Sigma}^3}\;\;.
\end{equation}

In view of the conservation of the total restmass $M_{\Sigma}$, we infer directly from
Eq.~(\ref{stationary1}) that ${\cal Q}_{\Sigma}$ evolves as:
\begin{equation}
\label{stationaryQ}
{\cal Q}_{\Sigma}=-\Lambda +
\frac{{\cal Q}_{\Sigma}(t_i)+\Lambda}{a_{\Sigma}^{3}}\;\;.
\end{equation}
For the same reason we infer from Eq.~(\ref{stationary2}) that $\gaverage{\CR}$
evolves as (inserting ${\cal C}^2$, Eq.~(\ref{constantC})):
\begin{equation}
\label{stationaryR}
\gaverage{\CR}=3\Lambda +\frac{\gaverage{\CR}(t_i)-3{\cal Q}_{\Sigma}(t_i)-6\Lambda}
{a_{\Sigma}^{2}} + \frac{3{\cal Q}_{\Sigma}(t_i)+3\Lambda}{a_{\Sigma}^{3}}\;\;.
\end{equation} 
The solution $\lbrace\,$(\ref{stationaryQ}), (\ref{stationaryR})$\,\rbrace$
satisfies the integrability condition (\ref{integrability}); it provides the first example of 
a non--trivial solution to the effective Einstein equations, 
in which kinematical backreaction and averaged scalar curvature are dynamically coupled. 
(This solution, for $\Lambda =0$ has been first discussed in connection with the 
Dark Energy problem in \cite{buchert:darkenergy}.)
 
This  solution stabilizes the (always positive--definite) combination
$6\Lambda + 3{\cal Q}_{\Sigma}-\gaverage{\CR}$, which evolves as
$(6\Lambda + 3{\cal Q}_{\Sigma}(t_i)-\gaverage{\CR}(t_i))/a_{\Sigma}^2 =
6 {\cal C}^2 /a_{\Sigma}^2 = 6 H_{\Sigma}^2$, so that for 
large $|a_{\Sigma}|$ it approaches zero. This relation governs the ``cross--talk''  
between kinematical backreaction, averaged scalar curvature and
$\Lambda$. Although the stationary state approaches the condition needed for a 
static state, the time--dependency of the individual terms changes the global picture 
drastically. For, if we prepare an initial state in the vicinity of the globally
static cosmos, the  condition (\ref{stationary1}) is conserved in time,
but the curvature evolves away from the condition  (\ref{stationary2}), 
which we shall explicitly discuss below for the subcase of a $\Lambda-$ free
stationary cosmos.

To work with the stationary models has an advantage: we can employ the dimensionless 
characteristics (\ref{omega}), which are valid on the global scale, and so 
ease the discussion of observational results. 
Directly from the stationarity conditions (\ref{stationary1})
and (\ref{stationary2}) we obtain for the dimensionless functionals:
\begin{equation}
\label{omegaS}
\Omega_m^{\Sigma} = - 4 \Omega_{\cal Q}^{\Sigma} + 2 \Omega_{\Lambda}^{\Sigma}\;\;\;
{\rm and}\;\;\;\Omega_{\cal R}^{\Sigma}=1+3\Omega_{\cal Q}^{\Sigma} -3 
\Omega_{\Lambda}^{\Sigma}\;\;.
\end{equation}
A straightforward calculation, employing the solution 
$\lbrace\,$(\ref{stationaryQ}), (\ref{stationaryR})$\,\rbrace$, provides their explicit
solutions in terms of two parameters -- chosen to be the mass density parameter at initial
time and the parameter for the cosmological constant at initial time: 
\begin{equation}
\label{parameters}
\Omega_m^i := \Omega_m^{\Sigma}(t_i) = \frac{2( {\cal Q}_{\Sigma}(t_i) + \Lambda)}
{3 {\cal C}^2} \;\;\;;\;\;\; 
\Omega_{\Lambda}^i := \Omega_{\Lambda}^{\Sigma}(t_i) = \frac{\Lambda}{3 {\cal C}^2} 
\;\;.
\end{equation}
Inserting the constant $\cal C$, Eq.~(\ref{constantC}), we get:
\begin{equation}
\label{solutionOmega}
\fl
\Omega_m^{\Sigma} = \frac{\Omega_m^i}{a_{\Sigma}}\;;\;\,
\Omega_{\cal Q}^{\Sigma} = -\frac{1}{4}\frac{\Omega_m^i}{a_{\Sigma}} + \frac{1}{2}
\Omega_{\Lambda}^i \,a_{\Sigma}^2 \;;\;\,
\Omega_{\cal R}^{\Sigma}=1 - \frac{3}{4}\frac{\Omega_m^i}{a_{\Sigma}} -\frac{3}{2}
\Omega_{\Lambda}^i \,a_{\Sigma}^2 \;;\;\, 
\Omega_{\Lambda}^{\Sigma}=\Omega_{\Lambda}^i \,a_{\Sigma}^2\;\,.
\end{equation}
The cosmic equation of state for a globally stationary cosmos
can be obtained
by inserting (\ref{stationary1}) and (\ref{stationary2})  
into (\ref{equationofstate}):
\begin{equation}
\label{cosmicstate2}
\fl
\frac{p_{\rm eff}}{\varrho_{\rm eff}} =: w_{\rm eff} \;=\; -\frac{1}{3} 
\left[\,\frac{1-3  \Omega_{\Lambda}^i \,a_{\Sigma}^2}
{1- \Omega_{\Lambda}^i \,a_{\Sigma}^2}\,\right]\;\;;\;\;
w_{\rm eff} (t_i) = -\frac{1}{3} \left[\, \frac{1}{1+ \frac{4\Lambda}
{3 {\cal Q}_{\Sigma}(t_i) -\gaverage{\CR}(t_i)}}\,\right]\;\;,
\end{equation}
which is time--dependent for a non--vanishing cosmological constant and, although 
it approaches, for large $|a_{\Sigma}|$, the cosmic state 
$p_{\rm eff} = - \varrho_{\rm eff}$, the asymptotic model is in a stationary 
and not in a de Sitter phase, since the cosmological constant is assumed to share the global 
balance condition. Below we shall assign another role to $\Lambda$ that will allow 
for a de Sitter phase too. For this purpose let us first drop $\Lambda$ as a possible source
sharing the stationarity condition.

\subsection{Globally  stationary cosmos without a cosmological constant}

Consider now again the subcase of a vanishing cosmological constant.
The cosmic equation of state for a globally stationary cosmos with $\Lambda =0$
can be inferred from Eq.~(\ref{cosmicstate2}):
\begin{equation}
\label{cosmicstate3}
{p}^{\Sigma}_{\rm eff}\;=\; 
-\frac{1}{3}\;\varrho^{\Sigma}_{\rm eff}\;\;.
\end{equation}
It is interesting to compare this condition with the investigation 
of backreaction in inhomogeneous cosmon fields by Christof Wetterich 
\cite{wetterich}, in particular with the  {\it cosmon equation of state}, which also points
to a possibly interesting different interpretation of the backreaction 
term ({\it cf.} Subsect.~\ref{subsect:inflation}). 
As already mentioned in the Introduction, 
in the case of a physical pressure source, the above equation of state 
allows for a static model without cosmological constant also in the homogeneous--isotropic
case (\cite{harrison:book}, p.383).
Recently, the cosmic equation of state in the present
formalism has been calculated in the framework of the Tolman--Bondi solution
\cite{nambu}, also confirming a strong coupling between averaged scalar curvature
and kinematical backreaction. For related discussions see \cite{celerier}, 
\cite{rasanen:LTB}, \cite{moffat}.

\noindent
The solution for this $\Lambda-$free cosmos, as a subcase of the general solution 
$\lbrace\,$(\ref{stationaryQ}), (\ref{stationaryR})$\,\rbrace$, reads:
\begin{equation}
\label{stationaryS}
{\cal Q}_{\Sigma} \;=\; \frac{{\cal Q}_{\Sigma}(t_i)}{a_{\Sigma}^{3}}\;\;\;;\;\;\;
\gaverage{\CR} \;=\;  \frac{\gaverage{\CR}(t_i)-3{\cal Q}_{\Sigma}(t_i)}{a_{\Sigma}^{2}} 
\;+\;\frac{3{\cal Q}_{\Sigma}(t_i)}{a_{\Sigma}^{3}}  \;\;.
\end{equation}
The total kinematical backreaction ${Q}_{\Sigma}V_{\Sigma} = 4\pi G M_{\Sigma}$ 
is a conserved quantity in this case. This solution 
implies for the dimensionless functionals:
\begin{equation}
\label{solutionOmegaS}
\Omega_m^{\Sigma} = \frac{\Omega_m^i}{a_{\Sigma}}\;\;;\;\; 
\Omega_{\cal Q}^{\Sigma} = -\frac{\Omega_m^i}{4 a_{\Sigma}}\;\;;\;\; 
\Omega_{\cal R}^{\Sigma}=1 - \frac{3\Omega_m^i}{4 a_{\Sigma}}\;\;;\;\; 
\Omega_{\Lambda}^{\Sigma}=0\;,
\end{equation}
where this cosmos only depends on a single parameter, which we have chosen to be
the restmass density parameter at the initial time $t_i$, as in Eq.~(\ref{parameters}).

As already mentioned in the general case, the averaged curvature evolves strongly in this cosmology.
In general, i.e. if the initial state is not the globally static state,  it evolves away from the
curvature of a static cosmos:  
\begin{equation}
\label{curvatureevolutionS}
\frac{\gaverage{\CR}}{12\pi G \gaverage{\varrho}}=1 - 
\frac{6{\cal C}^2}{12\pi G \gaverage{\varrho} a_{\Sigma}^2}=
1 - \left[\,1- \frac{\gaverage{\CR}(t_i)}{12\pi G \gaverage{\varrho}(t_i)}\,\right] 
a_{\Sigma}\;\;.
\end{equation}
Stating this remark in terms of the solution (\ref{stationaryR}) for $\Lambda =0$,
\begin{equation}
\label{curvatureevolutionR} 
\gaverage{\CR}=  \frac{3 {\cal Q}_{\Sigma}(t_i)}{a_{\Sigma}^3} - 
\frac{3{\cal Q}_{\Sigma}(t_i)-\gaverage{\CR}(t_i)}{a_{\Sigma}^2}\;\;,  
\end{equation}
where the numerators in both terms are positive--definite, we find that the second term
will dominate after some time, and the averaged curvature can change sign in the coarse of
evolution. In view of (\ref{solutionOmegaS}), this will happen 
when $\Omega_{\cal R}^{\Sigma}=1 - \frac{3\Omega_m^i}{4 a_{\Sigma}}=0$, i.e.  when the
density parameter has dropped to the ``critical'' value $\Omega_m^{\Sigma}=4/3$.

A comparison of this cosmology with a Friedmannian cosmology in terms of the evolution
of the single parameter $\Omega_m^i$ is performed in Appendix B.   

\subsection{Inflationary cosmogonies evolving into a globally static cosmos}
\label{subsect:inflation}

The inhomogeneous globally static cosmos and also stationary cosmologies were built on
the assumption that an exact balance
between averaged material mass and its kinematical fluctuations ({\it not} its 
density fluctuations) is established. 
The cosmological constant, if present, takes the role of a further player in an established globally
balanced state: if kinematical fluctuations vanish, then the Einstein static model
arises, if we have complete balance due to kinematical fluctuations, then $\Lambda =0$.
All intermediate states are conceivable.  

Now, instead of preparing such a non--accelerating state, we could further widen 
our model assumptions by allowing for a non--zero global acceleration.
To illustrate this we shall choose the simple assumption
that this acceleration is  entirely due to the cosmological constant, and also that
the corresponding expansion rate is also only due to $\Lambda$.  
In this case, the balance condition is established only among the
averaged material mass and the kinematical backreaction as in the globally static
model. We shall see that this ``driven balance'' (with the same cosmic equation of state
as in the static model) is 
conserved and the fluctuating matter is subjected to  a constant acceleration:
the conditions (\ref{static1}) and (\ref{static2}) are assumed to hold (for $\Lambda =0$), 
but the respective terms evolve in time according to the solution:
\begin{equation}
\label{mattercreation}
\gaverage{\CR}\;=\; \frac{\gaverage{\CR}(t_i)}{a^3_{\Sigma}}\;\;\;;\;\;\;{\cal Q}_{\Sigma} \;=\;
 \frac{{\cal Q}_{\Sigma} (t_i)}{a^3_{\Sigma}}\;\;\;;\;\;\;\gaverage{\varrho} =  
\frac{\gaverage{\varrho} (t_i)}{a^3_{\Sigma}}\;,
\end{equation} 
i.e., averaged scalar curvature  and kinematical backreaction both obey conservation
laws similar to the averaged density, and the ``driven balance'' is maintained.
With our assumptions, $a_{\Sigma}(t)$ is given by the exponential solution of the 
flat de Sitter model. (Note that $\gaverage{\CR}$ and ${\cal Q}_{\Sigma}$ with 
$\gaverage{\CR}=3{\cal Q}_{\Sigma}$
solve the integrability condition (\ref{integrability}) irrespective of
the particular form of $H_{\Sigma}={\dot a}_{\Sigma}/a_{\Sigma}$.)

The idea of inflation is related to matter creation. In this line
a global stationarity condition singles out a state 
which assumes that, as soon as matter is created, also its kinematical fluctuations 
are large and are trying to establish the balance condition on the global scale. 
To address this question properly,
the above model is too simple. A more ambitious model 
would first attempt to understand  kinematical 
backreaction and averaged curvature by means of effective inhomogeneous scalar fields.
We shall now discuss some aspects of such an attempt.

In the absence of matter and extrinsic curvature fluctuations,  
Eqs.~(\ref{averageraychaudhuri}),  (\ref{averagehamilton}) and 
(\ref{integrability}) imply (we always consider the global scale here):
\begin{equation}
\label{desitter}
3\frac{{\ddot a}_{\Sigma}}{a_{\Sigma}}  \;=\; \Lambda\;\;\;;\;\;\;
3\left( \frac{{\dot a}_{\Sigma}}{a_{\Sigma}}\right)^2  + \frac{\gaverage{\CR}(t_i)}
{2 a^2_{\Sigma}} 
 \;=\; \Lambda\;\;,
\end{equation}
leaving the exponentially expanding {\it de Sitter cosmos} as the only solution, 
when the initial scalar curvature is put to zero: $H_{\Lambda} = \sqrt{\Lambda /3}$.
As a next step, let us consider a cosmos without matter, but with intrinsic and 
extrinsic curvature fluctuations; the effective equations 
(\ref{averagehamilton}) and (\ref{integrability}) then read: 
\begin{equation}
\label{scalarfield1}
3\frac{{\ddot a}_{\Sigma}}{a_{\Sigma}} \;=\; \Lambda + {\cal Q}_{\Sigma}\;\;\;;\;\;\;
3\left( \frac{{\dot a}_{\Sigma}}{a_{\Sigma}}\right)^2 + \frac{\gaverage{\CR}+
{\cal Q}_{\Sigma}}{2} \;=\; \Lambda\;\;.
\end{equation}
As in (\ref{effectivefriedmann}) 
the above equations can be recast into standard zero--curvature Friedmann equations 
with new effective sources, now implying an interpretation in terms of an effective 
scalar field:
\begin{equation}
\fl
\varrho_{\Phi} :=\varrho_{\rm eff} = -\frac{1}{16\pi G}{\cal Q}_{\Sigma} - \frac{1}{16\pi G}
\gaverage{\CR}
\;\;;\;\;p_{\Phi}:=p_{\rm eff} = - \frac{1}{16\pi G}{\cal Q}_{\Sigma} + \frac{1}{48\pi G}
\gaverage{\CR}\;.
\end{equation}
In this matter--free case the above equations suggest an interpretation of the geometrical
degrees of freedom in the extrinsic curvature fluctuations as the kinetic energy 
part of a scalar field: ${\cal Q}_{\Sigma}$ obeys a {\it stiff} equation of state; also,
depending on the sign of ${\cal Q}_{\Sigma}$, we may have ``phantom energy''
\cite{caldwell}, (see also e.g. \cite{piazza}).
An important remark here: the simplified case of 
vanishing averaged scalar curvature implies through the integrability condition 
(\ref{integrability}) that ${\cal Q}_{\Sigma} = {\cal Q}_{\Sigma} (t_i) a_{\Sigma}^{-6}$, 
and the effective sources 
decay in proportion to the square of the inverse volume. As the previous 
considerations about globally stationary cosmologies have shown, the average
curvature plays a significant role for the maintainance of a large  ${\cal Q}_{\Sigma}$
and therefore, the presence of curvature is crucial and  
should not be neglected in this picture.

If we now take matter sources into account, the cosmic equation of state changes and it is
here, where detailed models of matter creation would have to be analyzed to understand
whether and if, physically, such a creation process would entail, e.g. strong global 
expansion fluctuations, favoring an inhomogeneous state at the time when $\Lambda$
no longer rules the expansion. If ${\cal Q}_{\Sigma}$ is interpreted as a kinetic
energy term of a scalar field (we may call it a {\it morphon}), then its very presence
would also shape the extrinsic curvature of created matter fluctuations; its
role in the equations above shows that a possible balance condition in the process
of conversion of kinetic energy into matter would be 
${\cal Q}_{\Sigma}=4\pi G \gaverage{\varrho}$, as required
for a globally static or stationary cosmos.

\subsection{Global far--from--equilibrium states}

Expressing a fluctuation--dominated cosmos in a thermodynamic language, the globally 
stationary state is in an {\it out--of--equilibrium state}
compared with the Friedmannian {\it ``equilibrium state''} (in the sense defined below). 
In \cite{hosoya:infoentropy} an entropy measure has been advanced that we can employ to
characterize both states: looking at
the non--commutativity relation (\ref{commutativity}) for $\Psi = \varrho\;$
on the global scale, Hosoya et al. \cite{hosoya:infoentropy} found that
the source of non--commutativity is given by the production
of {\it relative information entropy}, defined as to measure the 
deviations from the average mass density due to the development of inhomogeneities:
\begin{equation}
\label{commutationentropy}
\gaverage{\partial_t \varrho}-\partial_t \gaverage{\varrho}\; =\;
\frac{{\partial_t \,{\cal S}\lbrace\varrho || \gaverage{\varrho}\rbrace}}
{V_{\Sigma}}\;\;,
\end{equation}
with the, for positive--definite density, positive--definite {\it Kullback--Leibler functional}
\begin{equation}
\label{relativeentropy}
{\cal S} \lbrace \varrho || \gaverage{\varrho}\rbrace: \;=\; 
\int_{\Sigma} J d^3 X\;\varrho \ln \frac{\varrho}{\gaverage{\varrho}}\;\;.
\end{equation}
This measure vanishes for Friedmannian cosmologies (``zero structure'') 
and so defines the notion ``equilibrium state'' introduced above. 
It attains some positive time--dependent value otherwise.
The source in (\ref{commutationentropy}) shows that relative entropy production and
volume evolution are competing:  commutation can be reached, if the volume expansion
is faster than the production of information contained within the same volume.
 
In \cite{hosoya:infoentropy} the following conjecture was advanced:

\smallskip

{\sl The relative information entropy of a dust matter model
${\cal S} \lbrace \varrho || \gaverage{\varrho}\rbrace$
is, for sufficiently large times, globally (i.e. averaged over the
whole compact manifold $\Sigma$) an increasing function of time.}

\medskip

\noindent
This conjecture can be proven for linearized scalar perturbations on a Friedmannian 
background (the growing--mode
solution of the linear theory of gravitational instability implies $\partial_t \,{\cal S} >0$
and $\cal S$ is, in general, time--convex, i.e. $\partial^{2}/\partial_t^2 \,{\cal S} >0$).
However, in a generic non--perturbative situation it may not hold. 
Below we shall approach this question for the case of a globally stationary cosmos.

Let us first consider the general situation.
We calculate the second time--derivative of 
(\ref{relativeentropy}) and first obtain \cite{hosoya:infoentropy}:
\begin{equation}
\label{pdot}
\frac{{\ddot{\cal S}}}{V_{\Sigma}} \;=\;
-\gaverage{\delta\varrho\delta{\dot\theta}} + \gaverage{\varrho} 
(\Delta\theta)^2 \;\;,
\end{equation} 
where, for any scalar field $\Psi$,  $\delta \Psi := \Psi - \gaverage{\Psi}$ denotes the deviation 
from the global average value, and
$\Delta \Psi := \sqrt{\gaverage{\delta\Psi)^2}}$ the global amplitude
of its fluctuations.
Raychaudhuri's equation specifies the deviations $\delta{\dot\theta}$ for the time--evolution
of the expansion rate, 
\begin{equation}
\label{raychaudhuri}
\delta{\dot\theta} = - 4\pi G \delta\varrho - \frac{1}{3}\delta(\theta^2 ) - 
2 \delta(\sigma^2 )\;\;\;\;,\;\;\;\;\delta \Lambda =0\;\;,
\end{equation}
which, together with the {\em commutation rule} (\ref{commutativity})  yields:
\begin{equation}
\label{pdot1}
\frac{{\ddot{\cal S}}}{V_{\Sigma}} = 4\pi G(\Delta 
\varrho)^2+\gaverage{\varrho}(\Delta\theta)^2 + \frac{1} 
{3}\gaverage{\delta\varrho\delta\theta^2}+2\gaverage{\delta
\varrho\delta\sigma^2}\;\;.
\end{equation}
For our purpose the above equation may be recast by explicitly writing out the 
last second terms, and
expressing the expansion fluctuation amplitude in terms of 
${\cal Q}_{\Sigma} = \frac{2}{3}(\Delta\theta)^2 -2\gaverage{\sigma^2}$ 
({\it cf.} Eq.~(\ref{eq:Q-GR})). We 
obtain the general equation:
\begin{equation}
\label{pdot2}
\frac{{\ddot{\cal S}}}{V_{\Sigma}} = B_1 + \gaverage{\varrho}\left[\,{\cal Q}_{\Sigma} -
\frac{1}{3} \gaverage{\theta}^2 \,\right]\;\;,
\end{equation}
with $B_1 : = 4\pi G (\Delta\varrho)^2 + \frac{1}{3}\gaverage{\varrho\theta^2}+
2\gaverage{\varrho\sigma^2} \;\ge \;0$.

\smallskip

From this equation we infer that, for a globally static universe model 
with ${\cal Q}_{\Sigma}=4\pi G\gaverage{\varrho}$ and 
$\gaverage{\theta}=0$, the {\it Kullback--Leibler functional} is always 
time--convex. 

For the case of a stationary universe model
(with or without $\Lambda$) we can obtain sufficient conditions for time--convexity
as follows. 
Directly from (\ref{pdot2}) we conclude that ${\cal Q}_{\Sigma}-\frac{1}{3}
\gaverage{\theta}^2 = {\cal Q}_{\Sigma}-3 H_{\Sigma}^2$ should then be
positive and, employing the
dimensionless characteristics (\ref{omega}) on the global scale, that the condition
\begin{equation}
\label{sufficientQ}
\Omega^{\Sigma}_{\cal Q} \;\le\; - \frac{1}{2}
\end{equation}
is sufficient.
Inserting  the stationarity condition ${\cal Q}_{\Sigma}=4\pi G
\gaverage{\varrho} - \Lambda$ into (\ref{pdot2}) we instead need the condition
$4\pi G \gaverage{\varrho} - \Lambda -\frac{1}{3} \gaverage{\theta}^2 \;\ge \;0$ 
which, expressed through the average characteristics
(\ref{omega}) on the global scale, reads:
\begin{equation}
\label{sufficient2}
\frac{1}{2}\Omega^{\Sigma}_m - \Omega^{\Sigma}_{\Lambda}\;\ge\;1\;\;.
\end{equation}
For a $\Lambda-$free stationary cosmos we conclude that a sufficient condition for 
time--convexity of the Kullback--Leibler functional is $\Omega^{\Sigma}_m \ge 2$,
which implies a sufficiently positive global average curvature through Eq.~(\ref{omegaS}):
\begin{equation}
\label{sufficient3}
\Omega^{\Sigma}_{\cal R} = 1 - \frac{3}{4}\Omega^{\Sigma}_m \;\;\Rightarrow\;\;
\Omega^{\Sigma}_{\cal R} \;\le\;-\frac{1}{2}\;\;.
\end{equation}

Given the above results, we have three possibilities for the evolution of the
entropy functional (\ref{relativeentropy}) in a globally static cosmology (and also
in a globally stationary cosmology, if the conditions above are met), depending
on the sign of its first time--derivative, the relative entropy production rate
(Eq.~(\ref{commutationentropy}); note that for the special case of a globally static cosmos
$\partial_t\gaverage{\varrho} = 0$):

firstly, if ${\dot{\cal S}}$ is positive at initial time $t_i$, then the functional (\ref{relativeentropy})
is always growing; secondly, ${\dot{\cal S}}$ becomes positive at some time $t>t_i$, in
which case the functional (\ref{relativeentropy}) is always growing thereafter; thirdly,
${\dot{\cal S}}$ never becomes positive, i.e. the functional (\ref{relativeentropy}) 
approaches some constant ${\cal S}_0 \ge 0$ from above. This latter case would violate the
Entropy Conjecture advanced in \cite{hosoya:infoentropy}.
In the first two cases, looking back to Eq.~(\ref{commutationentropy}), 
the static model leaves the volume unchanged, while information is created. This 
enhances the source of non--commutativity, which may be interpreted as a signature of a 
non--Friedmannian state. 

The globally stationary cosmos does not necessarily imply a time--convex evolution of 
the relative information entropy. In this context it is interesting to think of 
a floating equilibrium 
realized in open biological systems (since gravity is long--ranged, 
a gravitational system is also not isolated). 
In this analogy a stationary far--from--equilibrium may be
characterized, according to Prigogine \cite{prigogine}, by a minimum of the entropy
production rate, i.e. for our inhomogeneity measure (\ref{relativeentropy}) we would 
require the condition: ${\ddot{\cal S}}\;=\;0$. Whether such a saturation 
is possible within a dust cosmology is doubtful in view of a lack of physics like 
pressure gradients. However, the above analogy prevails.

\subsection{Discussion: the global picture}

An investigation of models in the framework of the effective equations 
while still including a cosmological constant shows that the inflationary picture of 
cosmogonical scenarios in the Early Universe has not to be abandonned.
The cosmology can still undergo several phase transitions and inflationary phases. 
Like in the early days after Einstein's suggestion of the static cosmos, it is a subject of 
controversy related to the initial conditions of the universe model: either the singular
`Big Bang' initial state, or the static initial state eventually evolving into an expanding 
universe model is preferred (see the discussion in \cite{barrowetal:static}). 
In the context of {\it inhomogeneous} inflationary scenarios, we are entitled to
single out an averaged model that is globally stationary, i.e. 
we assume that inflation may end and, during a phase of damped oscillations,
exit into a globally stationary phase, instead of a homogeneous Friedmannian 
phase. If such an inhomogeneous ``initial condition'' at the end of inflation is 
prepared, then e.g.  a globally static inhomogeneous 
cosmos certainly is in a stronger position than his homogeneous predecessor with 
regard to its more general properties, especially if its stability could be proven.
Similar ideas have actually been already discussed within chaotic inflation, 
e.g.,\cite{linde1,linde2}, being mirrored here within a classical cosmos.

Justifying such an inhomogeneous initial state physically is another issue.
If we follow Einstein's thoughts leading to the static cosmos, the relevant arguments
were not empirical, but rather philosophical. These thoughts are far from 
conceiving a `fitting model' to observational facts. The {\it cosmological principle}
that elevates the Friedmannian models to a matter of principles is not always
the underlying reason to employ homogeneous--isotropic cosmologies. 
As the stationary but inhomogeneous cosmology shows, there is no 
local isotropy and, hence, the {\it cosmological principle} must be replaced by
another principle of ``global prejudice'' (in the constructive way of the word) 
like for example:  
``the Universe as a whole cannot create momentum out of itself''
(a ``non--M\"unchhausen'' principle\footnote{Baron v. M\"unchhausen was
trying to pull himself out of the pond without external help.}). 
The formulation of a sound principle is itself a considerable task;
the tentative balance condition that was helpful to construct exact examples,
should be replaced by a cosmic virial theorem averaged on the global scale, which in turn 
calls for a generalization of the dust matter model.  Also, global topological constraints
should be important. 
The way how to deal with a cosmological constant, e.g. the question whether we 
should include it in a virialized
state \cite{lahav:virial}, and the more fundamental approach
of an effective scalar field dynamics in the Early Universe is another issue of concern
in the formulation of a global principle. For example, a global net acceleration 
can be a result of initial 
conditions, and in fact this is the case in generic models of Friedmannian cosmology.

While the globally static cosmos features properties similar to his homogeneous
predecessor, e.g. the global density and the global curvature are constant in time,
the globally stationary cosmos evolves very differently. We have shown that,
although the  relation among kinematical backreaction and averaged curvature
tends to the corresponding relation in the globally static cosmos, the 
time--dependency of the individual terms 
may affect the global properties strongly. The stationary state tends to the
static state only in the sense that, e.g. in the case of an expanding cosmos, 
the rate of expansion slows down, but the steady increase of the scale factor allows for a global
change of the sign of curvature.   
As Eq.~(\ref{curvatureevolutionR}) shows, an initially
positive averaged scalar curvature would decrease, and eventually would become negative.
This may not necessarily be regarded as a signature of  a global topology change, as 
a corresponding sign change in a Friedmannian model would suggest; 
the averaged scalar curvature is only a weak descriptor for the topology in the general case,
and information on the sectional curvatures is required for definite conclusions on this
issue (see \cite{aubin} Ch.9; \cite{anderson}).
However, if a global change of topology arises, we could imagine that 
the cosmos started with 
a finite volume of a closed spaceform, then the averaged curvature of the 
inhomogeneous space may become negative and a globally hyperbolic 
curvature arises that would suggest a finite--volume compact spaceform with 
(now no longer simply--connected) sections of negative sectional curvatures and, generically,
a ``horned topology'' (for topology--related issues see 
\cite{uzan:topology}, \cite{lehoucq:topology}, \cite{aurich1:topology},
\cite{aurich2:topology}). 

\subsection{Discussion: the regional picture}

The implications that are furnished by the effective static model featuring 
{\it regional} expansion and contraction are qualitatively
very different from a cosmogony with {\it global} expansion: firstly, 
the former is more ``violent'' than the
latter, since fluctuations grow exponentially rather than in a moderate power--law pace;
secondly, an expanding region must be counterbalanced by a contracting region.
The second property implies that, depending on estimates of the total size of the cosmos,
we should be close to seeing a large ``blue excess'' in galaxy number counts, e.g. at the borders of
the currently drawn Sloan survey. The microwave sky would be redshifted in expanding domains 
alternating with a blue--shifted sky in contracting domains,
while we would have to claim an underdensity in our regional Hubble volume 
\cite{tomita1,tomita3}, \cite{wiltshire}, \cite{gronltb} 
counterbalanced by overdense regions beyond the horizon; 
we would be surrounded (in space 
and time direction) by high--density walls.
This spatially and temporally alternating scenario implies a large paradigmatic change
compared with the standard model of cosmology.
 
Exponential fluctuations imply the likely situation that, e.g. the global curvature
cannot be seen on the scale of our Hubble volume. Not unlikely, since we are
seeing a large Hubble expansion, the regional curvature is (relatively) more 
negative, i.e. in a special case seemingly zero or negative. 
As already noted, a globally static cosmos would necessarily call for a 
replacement and, regionally for an obvious refinement of the cosmological
principle including the question, 
whether and how close our observers have to be at the center of such a 
regional ``Hubble bubble'' \cite{tomita2}, \cite{linde2}, \cite{gronltb}. The scale of this ``reduced
curvature region''   likely exceeds scales that have been discussed in connection with
peculiar--velocity catalogues that are statistically affected by boundary conditions 
\cite{zehavietal}, \cite{giovanellietal}. A strong constraint is the isotropic 
microwave sky, a question related to assigning the whole dipol to our proper 
motion \cite{copelandetal}.

The globally stationary cosmology evolves very differently and, hence, also its
regional properties would be different. Since already the global curvature can
approach $0$, the expanding stationary cosmos may show similar regional 
fluctuations compared with a Friedmannian model, however, in a much 
narrower range of times. The differences to a Friedmannian model can be clearly
seen in the detailed comparison given in Appendix B.

We shall not embark here further into speculations about implications of a globally
static or stationary cosmos. It is clear that such a drastic change of views stimulates 
thoughts, but detailed calculations needed for proper statements do not exist yet.
Rather, we shall  put
a currently held discussion  about the Dark Energy problem into perspective, which turns out
to lie at the heart of the balance condition required for 
globally stationary cosmologies.

\section{The Dark Energy problem}

Let us first recall the contemporary view on the global universe model.
The {\it standard model of cosmology} idealizes the matter distribution in the
Universe to be homogeneous and isotropic, neglecting structure.
Observations point to an averaged matter content of at most 30 percent
(including dark matter) and almost flat space sections \cite{WMAP}.
The missing gap to fill the global cosmic triangle (\ref{triangle}) is modeled by a
(spatially constant) cosmological term in the simplest case \cite{peeblesratra}
(the sum of the parameters has to be equal to $1$). 
This is the {\it concordance model} (roughly $0.3 + 0.0 + 0.7 = 1$), which can be fitted to 
a large set of orthogonal observational data. 
(There are, however, other voices \cite{blanchardetal}, \cite{blanchard}.)
Modeling this dark energy gap with a cosmological constant results in an
accelerated phase\footnote{The only direct support for acceleration comes from
high--redshift supernovae observations \cite{schmidt04}; however, the `fitting models' for the interpolation
between low--and high--redshift observations and the measured distances do not
take inhomogeneities into account, see \cite{ellisbuchert} for a discussion
and references.}.
The `coincidence' that dark energy starts to dominate exactly when also
structure enters the non--linear regime suggests that there could be
a physical relation between the effect of structure on the average
expansion (known as {\it backreaction effect}) and the dark energy gap found in the
standard model, thus providing a natural solution to this {\it coincidence problem}. 
(It is clear here that sub--horizon fluctuations are made responsible.)

From Eq.~(\ref{averageraychaudhuri}) the condition for an accelerating patch $\CD$
of the Universe directly follows:
\begin{equation}
\label{accelerationcondition}
{\cal Q}_\CD \;>\;4\pi G \average{\varrho}\;\;.
\end{equation}
This regional condition is weaker than the requirement of global acceleration, since it 
accounts for the regional nature of our observations. 
It has been recently claimed \cite{kolbetal} that the condition 
(\ref{accelerationcondition}) could be satisfied 
within our regional Hubble volume, hence providing a smart explanation of
the Dark Energy problem \cite{wetterich}, \cite{beanetal:darkenergy}. 
In \cite{buchert:darkenergy} the following argument has been advanced.
The considerations in this paper and the ``classical'' explanation of the Dark Energy 
problem in terms of backreaction effects are intimately related through the condition
(\ref{accelerationcondition}).
If this condition would hold, then also the cosmologies presented in this work would attain
the status of physically viable models on the global scale, since the physical basis of
these models is exactly the possibility that kinematical expansion fluctuations 
and averaged sources are of the same order. A crucial support  for the condition 
(\ref{accelerationcondition}) to survive in time could also be given in
\cite{buchert:darkenergy} employing a particular example of the averaged Einstein
equations, listed as globally stationary cosmos without $\Lambda$ in this paper. 
This example of an exact solution  indeed points to
a strong coupling between kinematical backreaction and averaged scalar curvature:
the rate of decay of  ${Q}_{\Sigma}$ is in proportion to $\gaverage{\varrho}$ 
and can therefore be of the same order as $4\pi G\gaverage{\varrho}$ today.

\section*{Caveats} 

There is, however, a large body of opponents including myself who do not think that
the condition (\ref{accelerationcondition}) can be attained within the standard model
of cosmology. However, this opposition rests on the picture that fluctuations 
grow through gravitational instability of a homogeneous
universe model. If we allow for the very different picture of inhomogeneous initial
conditions, then the above outlined explanation can be a natural consequence.
In the standard picture of cosmological structure formation from Cold Dark Matter
initial conditions there are essentially three caveats that have to be overcome, which
we are going to put into perspective now in some more detail.

\subsection*{\sl Caveat 1: explicit calculations of kinematical backreaction}

The above summarized suggestion of solving the Dark Energy problem has a long history; there 
have been many attempts to calculate backreaction effects following the advent of 
the averaging problem initiated by George Ellis \cite{ellis}
(see \cite{ellisstoeger}, 
\cite{futamase1,futamase2,bildhauerfutamase1,bildhauerfutamase2},
\cite{russetal},\cite{brandenberger1,brandenberger2,brandenberger3}, \cite{wetterich},
\cite{rasanen:darkenergy}, \cite{kolbetal} and refs. therein).
The new input into the discussion concerns the quantitative importance of backreaction 
summarized by the ``bold claim'' (\ref{accelerationcondition}). Earlier efforts to support this
claim point to a negative answer. Let us therefore 
look into the details of explicit calculations of the effect. 

Cosmologists are mostly thinking in Newtonian terms when modeling
structures (e.g. by N--body simulations which are constructed such that
their spatial average evolves as in the standard model).
Computation of backreaction in the Newtonian framework is possible
in great detail \cite{bks}, but there is a drawback (often giving
rise to misunderstanding in the literature): the relevance of
kinematical backreaction for the global evolution cannot be estimated within
Newtonian cosmology, simply because one can in general prove that
an averaged Newtonian cosmology has zero global kinematical backreaction 
\cite{buchertehlers} ${\cal Q}_\CD$\footnote{The reason is that the 
kinematical backreaction term ${\cal Q}_\CD$ can be written as a
total divergence on Euclidean space sections,
and a Newtonian cosmology, in order to be {\it uniquely}
defined, has to impose periodic boundary conditions on the
inhomogeneities \cite{buchertehlers,ehlersbuchert}; 
hence, by Gauss' theorem, backreaction terms are just
boundary terms and they vanish for an empty boundary! 
Putting this fact into a constructive perspective, this non--trivial property of 
${\cal Q}_\CD$ proves
that the architecture of current N--body simulations is physically correct.}.
Nevertheless, as non--perturbative calculations in \cite{bks} show, backreaction
has indirect impact on the evolution of the standard cosmological
parameters on (relatively large) regional scales.
However, the dimensionless backreaction parameter (\ref{omega})
$\Omega_{\cal Q}^{\CD} := - \frac{{\cal Q}_{\CD}}{6 H_{\CD}^2 }$
on large {\it expanding} regions is quantitatively negligible compared with the 
effective density parameter
$\Omega_m^{\CD} : = \frac{8\pi G M_{\CD}}{3 V_{\initial{\CD}}a_{\CD}^3 
H_{\CD}^2 }$, i.e. the condition  (\ref{accelerationcondition}) rewritten in terms of 
dimensionless characteristics,
\begin{equation}
\label{accelerationconditionomega}
-\Omega_{\cal Q}^{\CD} \;>\;\frac{\Omega_m^{\CD}}{4}\;\;,
\end{equation}
(implying for the effective deceleration parameter (\ref{deceleration}) $q_{\rm eff}^\CD <0$)
is not fulfilled on large scales in these calculations. In fact, $\Omega_{\cal Q}^{\CD}$ is not only small,
but {\it positive}, i.e., with regard to (\ref{eq:Q-GR}), shear fluctuations dominate over expansion
fluctuations. In \cite{bks}
it was achieved to fulfill this condition on a patch of 
$100$ Mpc (for $h=0.5$) by saying that, on these
smaller scales, we may have different expectation values for velocity fluctuations 
(the kinematical components entering
${\cal Q}_\CD$ in Newtonian theory) 
and for the density fluctuations (they are theoretically
independent and, on $100$ Mpc, it may well happen that the expectation value for 
${\cal Q}_\CD$ is $1$--$\sigma$ off the expectation value for the density fluctuations, i.e. we 
could prepare an `untypical' sample). Although this situation is likely to happen (the probability
distribution of ${\cal Q}_\CD$ is non--Gaussian), we cannot apply this argument to very large samples.

To give another example (out of many that are not quoted here),
the answer attempted in the work by Russ et al. \cite{russetal} was based on a second--order
perturbative calculation in General Relativity. However, initial and boundary conditions were assumed 
such that, already from their assumptions, the global backreaction term, according to
our definition, has to vanish; see the discussion in \cite{buchert:grgdust}). 
Although it is obvious from the above remarks about 
Newtonian calculations that the global value can only be determined in the 
framework of General
Relativity, the quoted relativistic calculation was ``designed'' after a Newtonian model. This also
reveals  inherent difficulties for treating the problem with sufficient generality, if standard 
assumptions like periodic boundary conditions are adopted.

Given these remarks we could say that {\sl Caveat 1} is due to Newtonian or quasi--Newtonian 
work, or (since calculations have been mostly done perturbatively), is generally due to the fact that 
backreaction effects need a non--perturbative and background--free 
calculation (see \cite{langloisvernizzi} for suggestions of non--perturbative variables in 
General Relativity). Until more general calculations have been
done, we may say that the above quoted results cannot be extrapolated. However, basically 
a non--perturbative calculation in General Relativity has already been done: 
we have pointed out that Eq.~(\ref{averagefriedmann}) for the scale--factor is identical to the
corresponding Newtonian equation. The non--perturbative Newtonian model investigated in \cite{bks} 
was based on integrating this exact equation for a specific fluctuation source term, 
the (perturbative) `Zel'dovich approximation'.
Employing instead the relativistic version of this approximation as given by Kasai \cite{kasai} 
does not lead to 
differences concerning the results either.
Therefore, such a relativistic calculation for $a_\CD$ yields the same result in terms of 
${\cal Q}_\CD$. The difference is that the latter is linked to the scalar curvature (and
this may give rise to different interpretations as will be examined in a forthcoming work 
\cite{buchertetal:backreaction}). But, the quantitative conclusions given in \cite{bks} can to some extent
be extrapolated. This remark also points to the necessity of considering 
{\it initial conditions} 
other than those assumed in \cite{bks} in order to meet the condition 
(\ref{accelerationcondition}).

\subsection*{\sl Caveat 2: observational constraints}

The effective cosmological parameters defined in
(\ref{omega}) can be considered to provide a fair representation of the values which also an 
observer would measure in a sufficiently shallow survey region $\CD$ (the light--cone effect is not taken
into account). We may therefore discuss estimates of those parameters in comparison with
observed values.  Note, however, that the {\it interpretation} of observations is mostly done
by employing a standard Friedmannian cosmology as a `fitting model' and therefore, 
geometrical inhomogeneities (that are hidden in the definition of the spatial averages 
in the Riemannian volume element) are ignored; below we shall argue why we should not
ignore them.

On the grounds that the parameters (\ref{omega}) can be ``observed'', 
we examine the condition (\ref{accelerationconditionomega}), which itself would ``only'' require
a contribution of $|\Omega^\CD_{\cal Q} |>0,075$ for the value $\Omega_m^\CD =0,3$ 
to the cosmic quartet.
Assuming that this condition holds on some large
domain $\CD$, which we may take to be 
as large as our observable Universe, then Hamilton's constraint
in the form (\ref{hamiltonomega}) also implies: 
\begin{equation}
\label{curvaturecondition}
\Omega_{\Lambda}^{\CD} \;+\;\Omega_{\cal R}^{\CD}\;=\;
1\;-\;\Omega_m^{\CD}-\Omega_{\cal Q}^{\CD}
\;>\;1\;-\;\frac{3}{4}\,\Omega_m^{\CD}\;\;,
\end{equation}
showing that we need a substantial negative curvature (positive $\Omega_{\cal R}^{\CD}$)
on the domain $\CD$, if we put the 
cosmological constant to zero. (To reconcile this condition with a non--vanishing
cosmological constant would need
an even larger value of $\Omega_{\Lambda}^{\CD}$ than that suggested by the concordance model.)
The fact that a large value of kinematical backreaction entails a substantial average Ricci
curvature has also been stressed by  R\"as\"anen \cite{rasanen:constraints}.

The condition (\ref{curvaturecondition}) seems to contradict the widely agreed expectation that the
curvature should be very small, the concordance model assumes an exactly zero scalar curvature.
However, the Friedmannian curvature parameter is only directly related to the averaged scalar 
curvature in the homogeneous case.
If we would -- mistakenly -- estimate the averaged scalar curvature through 
$\Omega^\CD_k = - k_\CD /( a_\CD^2 H_\CD^2 )$, then we would 
have to estimate the (regional) time--integral of ${\cal Q}_\CD$ in the relation
$\Omega^\CD_{\Lambda}+\Omega^\CD_k \; =\;1\;-\;\Omega_m^{\CD}
-\Omega^\CD_{{\cal Q}N}$ ({\it cf.} Eq.~(\ref{omeganewton})), and thus our estimate would depend on the 
dynamical evolution of the kinematical backreaction parameter and not just on its value.
(In the (global) example presented in Appendix B we would have 
$\Omega^{\Sigma}_{k} = 1 - \Omega_m^i = const.$, 
while the true curvature functional $\Omega^{\Sigma}_{\cal R}$ is strongly time--dependent!)

If we ask whether the kinematical backreaction ${\cal Q}_\CD$ is observable, the answer within
our setup above is yes: on the observable domain $\CD$, ${\cal Q}_\CD$ is built from invariants
of the peculiar--velocity gradient in a Newtonian model. Ignoring again geometrical fluctuations
(and with them the fact that in a relativistic setting ${\cal Q}_\CD$ cannot be 
represented through invariants of a gradient, which is derived from a vector field) 
good high--resolution maps of 
peculiar--velocities could in principle determine the value of backreaction. Existing catalogues are,
however, too small and they would only return the cosmic variance 
around the assumed Friedmannian background in a likely untypical patch of the
Universe. 

On the other hand we know several observational facts that could place constraints
on the value of kinematical variables \cite{ellis:dahlem}.  
``Global'' bounds on ${\cal Q}_\CD$, where $\CD$ is of the order of the
CMB (Cosmic Microwave Background) scale, can be inferred from work of Maartens et al.
\cite{maartensetal1}, \cite{maartensetal2}:
observational upper bounds were given in terms of covariant and gauge
invariant quantities, e.g. for an upper bound on the shear fluctuations $\average{\sigma^2}$ we 
use the limit $\sigma^2_* / \theta^2_*\;<\;16\alpha^2\,10^{-10}$, 
where both $\sigma_*$ and $\theta_*$ have to 
be interpreted as averaged values in our setting, i.e. $\sigma_* := \average{\sigma}$ and 
$\theta_* := 3 H_\CD$; the parameter $\alpha$ is determined by 
observations and depends on scale; for large--scale CMB observations we set 
$\alpha \approx 1$.  
In the case when shear fluctuations would 
dominate, we may assume $\average{\theta}^2 = \average{\theta^2}$.
In that case, ${\cal Q}_\CD$ reduces to 
$-2 \average{\sigma^2}$ (is therefore negative); the corresponding dimensionless 
parameter obeys the bound
\begin{equation}
\label{bound1}
0\;<\; \frac{-{\cal Q}_\CD }{6H^2_\CD }= \Omega_{\cal Q}^\CD \;<\;2,4 \,\alpha\;10^{-9}  
\;\;\;{\rm for}\;\;\;\average{\theta}^2=\average{\theta^2}\;\;.
\end{equation}
This bound is comfortably satisfied in the calculations quoted above, but it is not what we 
need for explaining Dark Energy: the opposite situation must be examined, namely, we assume
shear fluctuations to be negligible and search bounds for a positive ${\cal Q}_\CD \approx 
\frac{2}{3} \left(\average{\theta^2} - \average{\theta}^2 \right)$.

On the Hubble length $L_H = 2c / H \approx 6 Gpc/h$ Maartens and his collegues
\cite{maartensetal2} obtain an upper bound
on the density fluctuations that we may convert into expansion fluctuations using,
for simplicity, the proportionality
relation stemming from the standard linear theory of gravitational instability on 
an Einstein--de Sitter reference background with density $\varrho_H$ and Hubble expansion
parameter $H$ (suppressing the decaying mode):
\begin{equation}
\label{lineartheoryfluctuations}
\frac{\average{\varrho - \average{\varrho}}^2}{\varrho_H^2}\;=\;
\frac{\average{\theta - \average{\theta}}^2}{V_i^2 H^2}\;\;,
\end{equation}
where the initial volume $V_i =4/3 \pi (L_H /2)^3$ accounts for the dimension and the scale.
With these assumptions and the upper bound on the density fluctuations 
quoted in \cite{maartensetal2}, the backreaction parameter is bound from below:
\begin{equation}
\label{bound2}
\fl
0\;>\; \frac{-{\cal Q}_\CD }{6H^2_\CD }= \Omega_{\cal Q}^\CD \;>\;
-\left(\,\frac{14}{5\Omega^\CD_{m}} + \frac{1}{10}\,\right)^2\; \alpha^2\;10^{-8}  
\;\;\;{\rm for}\;\;\;\average{\sigma^2 } =0\;,
\end{equation}
i.e., for $\Omega^\CD_{m}=0,3$ 
on the Hubble--scale we would have $\Omega_{\cal Q}^\CD \;>\; - 8,9 \,\alpha^2\;10^{-7}$.

However, a precise relation of the bounds on gauge--invariant variables 
given by Maartens \& collegues  with the relevant variables discussed here 
requires more efforts and is the subject of a forthcoming work.
Notwithstanding, the above 
order--of--magnitude estimates demonstrate that observational constraints will also 
provide obstacles to advocate the condition (\ref{accelerationcondition}).

The loophole here: the strong coupling of density fluctuations to expansion 
fluctuations, as given by a direct proportionality for small perturbations of a 
Friedmannian model, Eq.~(\ref{lineartheoryfluctuations}), was strongly used
to place constraints on expansion fluctuations. In an out--of--equilibrium state
these two types of fluctuations (being  independent variables) are decoupled and
an observational constraint must be based directly on an upper bound on
expansion fluctuations. 

\subsection*{\sl Caveat 3: geometrical fluctuations, curvature backreaction and the choice of 
foliation}

It is a reasonable question to ask whether a cosmological model is accelerating or not
given the framework of the kinematically averaged Einstein equations, provided we 
consider the chosen foliation of spacetime appropriate. Caveat 3 consists of 
the fact that those averages are performed on an inhomogeneous hypersurface.
This is all right as long as model and observations are to be compared, since observations
themselves are carried out in an inhomogeneous geometry. The problem we speak about
here comes into the fore by `fitting' a Friedmannian cosmology to observational data,
and this is the standard procedure. The question whether a cosmological constant
is needed to explain the observations is due to this fitting process. 
Consequently, there is a need to compare averages of the dynamical model to 
averages on a homogeneous geometry, which is a difficult task because the averaging
of tensors is not straightforward \cite{ellis,ellisstoeger,ellisbuchert}.
There is obviously an interpretation problem in
the standard model (having Euclidean geometry in the case of the
concordance model): averages on inhomogeneous spaces have not the
same values as averages on homogeneous spaces. 
Riemannian volumes may substantially differ from Euclidean
volumes and are not small perturbations {\it per se} (a Riemannian ball has $\pi^2 /6$
less volume than an Euclidean ball with the same geodesic 
radius \cite{buchertcarfora3}).
It is possible to define a
``geometrical renormalization'' flowing averages on inhomogeneous spaces
into Friedmannian averages \cite{klingon}. The consequences for
the regional cosmological parameters are summarized in \cite{buchertcarfora:PRL}
implying further ``backreaction effects'' such as {\it Ricci curvature backreaction} that
lead to a ``dressing'' of cosmological parameters by smoothed--out geometrical
inhomogeneities. These effects may provide a further key for explaining Dark Matter
and Dark Energy.
 
The answer to the question whether these effects are significant
needs detailed realistic models for ``volume roughening''
of spatial slices in the Universe, and robust estimators for intrinsic curvature
fluctuations. Since this problem has
been largely overlooked when considering averages in cosmology, there
are not much results. A naive swiss--cheese model by just glueing Riemannian balls
in place of Euclidean balls in the slice yields an effect of 67\%
reducing the necessary dark energy roughly from 70\% to 50\% \cite{buchertcarfora3}.
A mismatch of similar magnitude has been reported by Hellaby \cite{hellaby} using 
volume--maching as suggested by Ellis \& Stoeger \cite{ellisstoeger} (compare 
\cite{tanimoto}).
However, the quantitative significance of these effects on cosmological scales 
have yet to be explored.

Related to the averaging of both extrinsic and intrinsic inhomogeneities 
Zalaletdinov \cite{zal97} proposed a macroscopic
description of gravitation based on a covariant spacetime averaging
procedure. The geometry of the macroscopic spacetime follows from averaging
Cartan's structure equations, leading to a definition of correlation
tensors. Macroscopic field equations (averaged Einstein equations) can be
derived in this framework. Within this approach Coley et al. \cite{zala05} have 
recently shown that for a spatially homogeneous and isotropic macroscopic
spacetime, the correlation tensor is of the form of a spatial
curvature term. In this context it was also speculated that the extra correlation terms
might help to stabilize a globally static universe model.

As for any averaging procedure the choice of spatial slices on which one considers average
quantities is crucial. In \cite{klingon} this problem has been identified and, in the framework
of smoothing the geometry with the help of the Ricci flow, one has strong means to 
approach the slicing problem, e.g. by aiming at ``minimizing'' artificial
gauge effects  \cite{klingon}. 
Recently, Ishibashi and Wald \cite{wald} gave particular examples that show why 
caution is in order concerning a straightforward observational interpretation of averaged 
kinematical quantities on given spatial slices.
Here, a clear advantage of volume--averaging scalar quantities is that the resulting averages
of 4--covariant variables and their governing equations can be transformed to a different 
choice of slicing, which is a feature of the covariant fluid gauge \cite{ellisbruni} underlying the 
averaged equations in the present paper; examples that illustrate this 
have been given in \cite{buchert:grgfluid}. 

\section{Concluding Remarks}
\label{conclusion}

In this work we have contrasted Friedmannian cosmology to particular choices of
inhomogeneous cosmologies. Two very different points of view concerning the 
comparison of an idealized Friedmannian cosmos and an averaged inhomogeneous cosmos
are conceivable: 

Firstly, an averaged inhomogeneous model could be a small
perturbation of a Friedmannian model in the sense that the kinematical backreaction on
the global scale, ${\cal Q}_{\Sigma}$, is small and, hence, the kinematics of both models
measured by the evolution of the effective scale factor $a_{\Sigma}$ are comparable.
Still, as shown in detail within Newtonian cosmology in \cite{bks}, this must not imply
that the parameters of a Friedmannian model evolve as in the inhomogeneous model;
on the contrary, even for negligible backreaction, there are significant differences.
We may place such a universe model into a {\it near--equilibrium state} as measured, e.g.
by the information--theoretical measure (\ref{relativeentropy}) that vanishes for a
Friedmannian cosmology. 

Secondly, and this point of view was exploited in the present work, we may conceive
a universe model in a {\it far--from--equilibrium state}, characterized by strong
averaged expansion fluctuations on the global scale. Such a model is no longer a perturbation
of a Friedmannian model. We have shown that these highly--inhomogeneous models may 
then evolve in the vicinity of the balance condition ${\cal Q}_{\Sigma} = 
4\pi G \gaverage{\varrho}$ (disregarding the cosmological constant).

Which of these two points of view is realized depends on the choice of the initial state
of the Universe at the exit epoch of an eventual inflationary phase in the Early Universe.
We have argued why a possible solution of the {\it Dark Energy problem} through
kinematical backreaction effects is possible from the second point of view, but unlikely from
the first point of view.
Given such a global state we have shown that strong kinematical fluctuations are conserved, 
and the state is maintained by a strong coupling of backreaction to the averaged 
Ricci curvature. This has been exemplified by an exact solution of the averaged
Einstein equations for a globally stationary cosmos.
We may say that large kinematical fluctuations are maintained on the cost of averaged
scalar curvature in this solution. 

We have considered three families of cosmologies that belong to the out--of--equilibrium
state of vanishing global acceleration: the first family covers globally stationary cosmologies
without a cosmological constant; as a subcase, a globally static cosmos is possible
that revives the original Einstein cosmos concerning its global properties, however,
having very different properties on regional scales. The second family covers models
that obey the stationarity condition of vanishing global acceleration, but include
a cosmological constant sharing the balance condition. 
In this case we have also given the exact solution. Finally, the
third family rests on the stationarity conditions of the first family, but the global
acceleration is non--zero due to the cosmological constant alone. Fluctuations and
average characteristics in such $\Lambda-$driven cosmologies have very similar 
properties to the previous cosmologies, however, with the presence of a global
exponential expansion.
 
Although, formally, these new families of cosmologies obey simple equations and are
in this respect very close to Friedmannian cosmologies, their dynamical properties must
be regarded as significantly different. Investigating details of the new models must be
considered as an endeavour of high magnitude, especially if one is interested in the
regional properties. One important reason for
this difficulty lies in the scale--dependence of the average characteristics. 
For example, the Einstein static model has globally and locally a constant curvature 
and a constant density, while the globally static, but inhomogeneous cosmos features
strong matter and curvature fluctuations on regional scales. The latter
model enjoys the freedom that the averages can be scale--dependent, e.g., the Einstein
static model has (like all Friedmannian cosmologies) a scale--independent density, while
the globally static cosmos allows for a decay of the average density by going to larger
scales. It is possible that fluctuations, encoded in ${\cal Q}_{\CD}$, on a regional domain
$\CD$  are large relative to the averaged density $4\pi G \average{\varrho}$ 
on that domain,
but  ${\cal Q}_{\Sigma}$ could be very small on the global scale 
$|\Sigma|$ in absolute terms,
and still of the order of $4\pi G \gaverage{\varrho}$.

These new models need detailed investigation before any conclusive result 
relevant to observations can be obtained. In light of the present investigation three
lines of research appear to be fruitful strategies: firstly, 
investigating global stability properties of the presented cosmologies, i.e. studying 
perturbations on non--Friedmannian ``backgrounds'' including studies of
gravitational instability on regional domains; secondly, investigating
the relativistic generalization of \cite{bks} and comparing standard with
globally inhomogeneous initial conditions \cite{buchertetal:backreaction} 
and, thirdly, investigating inhomogeneous
inflationary scenarios and the properties of fluctuations at the exit epoch, i.e. studying
the role of kinematical backreaction in the Early Universe. 

\vspace{5pt}

\noindent
{Acknowledgements:}

\noindent
{\footnotesize
This work was supported by the Sonderforschungsbereich SFB 375 
`Astroparticle physics' by the German science foundation DFG.
Special thanks go to Mauro Carfora and Ruth Durrer for valuable remarks on the manuscript,
and for their hospitality and support during visits to the University of Pavia, Italy, and to the 
University of Geneva, Switzerland, as well as to Sabino Matarrese and Syksy R\"as\"anen for 
interesting discussions.}

\renewcommand{\theequation}{A.\arabic{equation}}
\setcounter{equation}{1}  
\section*{Appendix A: the size of a globally static cosmos}

In order to get an idea about typical numbers characterizing  a globally static cosmos, 
we are going to discuss some numerical estimates. 
We already imply that
there exist regional fluctuations of this global state; we take the classical Einstein cosmos as
the global model and infer our estimates from regionally observed properties. 

Dividing Friedmann's 
differential equation (\ref{friedmann2}) by $H^2$ (here, the static case is excluded),
we obtain Hamilton's constraint 
for the homogeneous--isotropic case, written in the iconized form of 
a {\it cosmic triangle} \cite{bahcall}:
\begin{equation}
\label{triangle}
\fl \quad
\Omega_m \,+\, \Omega_{\Lambda}\,+\,\Omega_k \;=\; 1\;\;;\;\;
\Omega_m := \frac{8\pi G \varrho_H}{3 H^2}\;\;,\;\;
\Omega_{\Lambda}:=\frac{\Lambda}{3 H^2}\;\;,\;\;
\Omega_k := \frac{-k}{a^2 H^2}\;\;.
\end{equation}
Confining ourselves now to our regional Hubble volume, we write 
$\Omega_m ({\rm\scriptstyle Hubble-scale})=: \alpha_m$ and express $H$ through $h := H / 100 Km/sMpc$. 
From (\ref{einstein3}) with $k=+1$ and (\ref{einstein4}) we first obtain:
\begin{equation}
\label{blackhole}
a_E = \frac{1}{\sqrt{4\pi G \varrho_E}}\;=\;\frac{1}{\pi} 2 G M_E = \frac{1}{\pi}
a_{\rm Schwarzschild}\;\;.
\end{equation}
Extrapolating the value of the
matter density $\varrho_H = \frac{3}{8\pi G}\alpha_m H^2 = \varrho_E$, we find for
the Einstein radius in space units:
\begin{equation}
\label{einsteinradius}
r_E = c a_E =  \frac{c}{\sqrt{\alpha_m}h} \sqrt{2/3}\;10^{-8} \frac{s Gpc}{m} \;\;,
\end{equation}
which, for $\alpha_m \approx 0,3$ and $h \approx 0,6$ yields $r_E \approx 7,5 \,Gpc$.
For comparison:
if $\alpha_m =1$ and $h \approx 0,46$ \cite{blanchardetal} we would get 
$r_E \approx 5,3 \,Gpc$, for  $\alpha_m \approx 0,3$ and $h \approx 0,46$, 
$r_E \approx 9,7 \,Gpc$. Defining a
{\it Hubble volume} roughly by $V_H = \frac{4\pi}{3} (c/H)^3$ (where we
took the horizon radius of a Friedmannian model, $r_H:=c/H \approx 3 Gpc/h$, 
as radius of an Euclidean sphere randomly placed within the Einstein cosmos), 
we get for the first set of  values 
above $V_H \approx 174,5 \,Gpc^3$, and with 
$V_E = 2\pi^2 r_E^3 \approx 8319 \,Gpc^3$, we  conclude that a 
fluctuating static cosmos with the radius of the homogeneous Einstein cosmos 
would roughly contain $50$ Hubble volumes.
Thus, the total volume $V_E$  is comfortably large to allow for significant 
regional fluctuations on the Hubble scale.
 
The regional curvature that we would observe on the Hubble scale will not be the 
global curvature of the Einstein cosmos. However, if we assume for simplicity 
that the curvature is non--fluctuating, 
but that we have a fluctuating expansion being globally zero
and regionally given by $h$ -- as will be possible in the general model discussed
in the text), we could estimate the global {\it Friedmannian curvature parameter} 
from regional observations: setting 
$\Omega_k ({\rm\scriptstyle Hubble-scale})
= :\alpha_k = -k /  a_E^2 H^2$, we obtain with the first set of values above
$\alpha_k = -\frac{c^2}{r_E^2 h^2} \,10^{-16} s^2 Gpc^2 /m^2 \approx - 0,44$. 

Thus, if we would `fit' a Friedmannian model within the regional Hubble volume, 
we would obtain
from (\ref{triangle}) $(0,3; -0,44; 1,14)$ for values of the regional cosmic triangle 
$(\alpha_m; \alpha_k; \alpha_{\Lambda})$, i.e. straight application of this (quickly cooked) 
model would be in trouble compared with the values of the so--called 
{\it concordance model} as a `best--fit'
Friedmannian cosmology in comparison with observations, $(0,3; \pm 0; 0,7)$.

However, these estimates are naive in a variety of ways: firstly, the averaged curvature is
{\it not} given by the Friedmannian curvature parameter, if kinematical 
fluctuations are present; secondly, a fluctuating cosmos implies fluctuations in
scalar curvature too, i.e. it is easily conceivable that the magnitude of the 
regional curvature could be smaller than the above estimate. 
Moreover, as we have learned, if expansion fluctuations are assumed large, then
we can as well dismiss the cosmological constant altogether; thirdly, 
a further feature of globally static inhomogeneous cosmological 
models is that they would allow for a significantly larger size compared to the values quoted
above, since the average density may likely be
scale--dependent and may decrease towards the global scale. 

\renewcommand{\theequation}{B.\arabic{equation}}
\setcounter{equation}{1}  
\section*{Appendix B: evolution of $\boldsymbol{\Omega}$--parameters in 
the globally stationary cosmos compared with their evolution in 
Friedmannian cosmologies for $\boldsymbol{\Lambda}{\bf=0}$}

At first glance, the reader may think that a globally 
stationary cosmos, introduced in Sect.~3, looks very similar to a 
Friedmannian cosmos, since the scale factor has a powerlaw form,
$a_{\Sigma} \propto t$, and the global expansion $H_{\Sigma} \propto t^{-1}$, but in fact
both cosmologies are drastically different in nature. This can be nicely illustrated with
the evolution of the density parameter, which is the only free parameter in both cosmologies,
if $\Lambda =0$. 

For this comparison 
we employ Eq.~(\ref{averagefriedmann}) on the global scale, 
and evaluate this equation with the solution
of the stationary cosmology ${\cal Q}_{\Sigma} = {\cal Q}_{\Sigma}(t_i) a_{\Sigma}^{-3}$:
\begin{equation}
\label{averagefriedmannS}
\qquad\fl
H_{\Sigma}^2 (1-\Omega^{\Sigma}_m ) a_{\Sigma}^2 + k_{\Sigma} =\frac{2}{3}
\int_{t_i}^t dt' \; {\cal Q}_{\Sigma} a_{\Sigma}' a_{\Sigma} = -\frac{2}{3}  
{\cal Q}_{\Sigma} (t_i) \left( \frac{1}{a_{\Sigma}} -1 \right)\;\;,
\end{equation}
with $k_{\Sigma} = ( {\cal Q}_{\Sigma} (t_i) + \gaverage{\CR} (t_i) )/6
= H_i^2 (\Omega^i_m -1)$, and 
the density parameter $\Omega^{\Sigma}_m$ in the general model (\ref{omega}). 
From the  definition of the latter we have a second equation:
\begin{equation}
\label{omegaevolutionA}
\Omega^{\Sigma}_m = \Omega^i_m \frac{H_i^2}{H_{\Sigma}^2 a_{\Sigma}^3}\;\;.
\end{equation}
Inserting (\ref{omegaevolutionA}) into (\ref{averagefriedmannS}), a simple calculation
provides:
\begin{equation}
\label{omegaevolutionB}
\Omega^{\Sigma}_m = \frac{\Omega^i_m}{\Omega^i_m (1-\kappa) + 
\left[\, 1- \Omega^i_m (1-\kappa)  \,\right] a_{\Sigma}}\;\;,
\end{equation}
with $\kappa = 1$ for the stationary inhomogeneous model with non--zero 
kinematical backreaction, ${\cal Q}_{\Sigma} (t_i) = \frac{3}{2}\Omega^i_m H_i^2$, 
and $\kappa =0$ for the Friedmannian models. 
This minor formal difference amounts to a substantially different evolution; the reader 
may insert numbers to compare the following solutions:
\begin{equation}
\label{omegaevolutionC}
\qquad\fl
\Omega_m ({\rm\scriptstyle Friedmann}) = 
\frac{\Omega^i_m}{\Omega^i_m + (1- \Omega^i_m) a_{\Sigma}}
\;\;\;;\;\;\;\Omega_m^{\Sigma} ({\rm\scriptstyle Stationary}) = \frac{\Omega^i_m}{a_{\Sigma}}
\;\;.
\end{equation}
Introducing the redshift $z$ in the Friedmannian model, 
$a_{\Sigma} = \frac{1+ z_0}{1 + z}$, and accordingly a (formal) effective redshift 
$z^{\rm eff}$ in the inhomogeneous model, we can write the above
solutions in terms of the density parameter {\it today}, $\bar{\Omega}^i_m := 
\Omega^{\Sigma}_m (z=0)$, as follows:
\begin{equation}
\label{omegaevolutionD}
\qquad\fl
\Omega_m ({\rm\scriptstyle Friedmann}) = 
\frac{\bar{\Omega}^i_m (1+z)}{1+\bar{\Omega}^i_m z}
\;\;\;;\;\;\;\Omega_m^{\Sigma} ({\rm\scriptstyle Stationary}) = \bar{\Omega}^i_m 
(1+ z^{\rm eff})\;\;.
\end{equation}
The scalar curvature in a Friedmannian cosmology, $\Omega_k = 1 - \Omega_m$ 
cannot change sign, but the averaged scalar curvature in the globally stationary cosmos,
$\Omega^{\Sigma}_{\cal R} = 1 - \frac{3}{4} \Omega^{\Sigma}_m$
(see Eq.~(\ref{solutionOmegaS})) can.
(Note, that the ``Newtonian'' curvature functional, introduced in Eq.~(\ref{omeganewton}),
stays constant in the stationary solution: $\Omega_k^{\Sigma} = 1-\Omega^i_m$!)
 
While the density parameter  
in a Friedmannian model stays smaller (larger) 
than $1$, if it is initially smaller (larger) than $1$, and while the curvature parameter 
stays smaller or larger than $0$ determined by initial data, 
the corresponding parameters (and the additional backreaction parameter) in the
stationary model can ``communicate with each other'' and can 
``freely operate'' (of course, subjected to the Hamiltonian constraint, 
$\Omega^{\Sigma}_m +\Omega^{\Sigma}_{\cal R}+\Omega^{\Sigma}_{\cal Q}=1$,
and the stationarity conditions) in the range of values:
\begin{equation}
\label{range}
\qquad\fl
+\infty \;>\; \Omega^{\Sigma}_m \;>\;+\frac{4}{5}\;\;;\;\;\;\;
-\infty \;<\; \Omega^{\Sigma}_{\cal R} \;<\;+\frac{2}{5}\;\;;\;\;\;\;
-\infty \;<\; \Omega^{\Sigma}_{\cal Q} \;<\;-\frac{1}{5}\;.
\end{equation} 
(These conditions can be easily derived from the fact that 
$3 {\cal Q}_{\Sigma} + \gaverage{\CR} = 6 H_{\Sigma}^2 \;>\;0$.)
The infinities are reached at the ``Big Crunch'' of a contracting stationary model
with ${\cal C}\;<\;0$, in which case $a_{\Sigma}\;\rightarrow\;0$ in a finite time.

It should be emphasized that these are the global values of the cosmological parameters; 
on regional scales very different values may be measured, whereas in the standard model 
of cosmology global and regional values have to be equal.


\section*{References}

\begin{thebibliography}{2005}

\bibitem{alametal}
U. Alam, V. Sahni, T.D. Saini and A.A. Starobinskii, Mon. Not. Roy. Astro. Soc. {\bf 344}, 
1057 (2003).

\bibitem{gronltb}
H. Alnes, M. Amarzguioui and {\O}. Gr{\o}n, astro--ph/0512006 (2005).

\bibitem{anderson}
M. T. Anderson, {\it Comparison Geometry}, MSRI Publications {\bf 30}, Berkeley CA, 
pp.49-82 (1997).

\bibitem{aubin}
T. Aubin, {\it Some Nonlinear Problems in Riemannian Geometry}, Springer Monographs
in Mathematics (1998).

\bibitem{aurich1:topology}
R. Aurich and F. Steiner, Mon. Not. Roy. Astro. Soc. {\bf 323}, 1016 (2001).

\bibitem{aurich2:topology}
R. Aurich and F. Steiner, Mon. Not. Roy. Astro. Soc. {\bf 334}, 735 (2002).

\bibitem{bahcall}
N. Bahcall, J.~P. Ostriker, S. Perlmutter and P.~J. Steinhardt, Science {\bf 284},  1481  (1999).

\bibitem{barrowetal:static}
J.D. Barrow, G.F.R. Ellis, R. Maartens and C.G. Tsagas, Class. Quant. Grav. {\bf 20}, L155 (2003).

\bibitem{beanetal:darkenergy}
R. Bean, S. Carroll and M. Trodden, astro--ph/0510059 (2005).

\bibitem{bildhauerfutamase1}
S. Bildhauer and T. Futamase, Mon. Not. Roy. Astro. Soc. {\bf 249}, 126 (1991).

\bibitem{bildhauerfutamase2}
S. Bildhauer and T. Futamase, Gen. Rel. Grav. {\bf 23}, 1251 (1991).

\bibitem{blanchard}
A. Blanchard,  Astrophys. \& Space Sci. {\bf 290}, 135 (2004).

\bibitem{blanchardetal}
A. Blanchard, M. Douspis, M. Rowan--Robinson and S. Sarkar, Astron.\ Astrophys.
{\bf 412}, 35 (2003).

\bibitem{brandenberger1}
R.H. Brandenberger, hep--th/0210165 (2002).

\bibitem{buchert:grgdust}
T. Buchert, Gen. Rel. Grav. {\bf 32}, 105 (2000).

\bibitem{buchert:grgfluid}
T. Buchert, Gen. Rel. Grav. {\bf 33}, 1381 (2001).

\bibitem{buchert:jgrg}
T. Buchert, in: {\em 9th JGRG Meeting, 
Hiroshima 1999}, Y. Eriguchi et al. (eds.), pp. 306--321 (2000); gr-qc/0001056.

\bibitem{buchert:darkenergy}
T. Buchert, Class. Quant. Grav. {\bf 22}, L113 (2005).

\bibitem{klingon}
T. Buchert and M. Carfora, Class. Quant. Grav. {\bf 19}, 6109 (2002).

\bibitem{buchertcarfora:PRL}
T. Buchert and M. Carfora, Phys. Rev. Lett. {\bf 90}, 031101-1-4 (2003).

\bibitem{buchertcarfora3}
T. Buchert and M. Carfora, in: {\it 12th JGRG Meeting},
Tokyo 2002, M. Shibata et al. (eds.), 157--61 (2003); astro-ph/0312621.

\bibitem{buchertehlers}
T. Buchert and J. Ehlers, Astron. Astrophys. {\bf 320}, 1 (1997).

\bibitem{bks}
T. Buchert, M. Kerscher and C. Sicka, Phys. Rev. D. {\bf 62}, 043525 (2000).

\bibitem{buchertetal:backreaction}
T. Buchert et al., in preparation.

\bibitem{caldwell}
R.R. Caldwell, Phys. Lett. B {\bf 545}, 23 (2002).

\bibitem{celerier}
M.--N. C\'el\'erier, Astron. Astrophys. {\bf 353}, 63 (2000).

\bibitem{zala05}
A.A. Coley, N. Pelavas and R.M. Zalaletdinov, 
Phys. Rev. Lett. {\bf 95}, 151102 (2005).

\bibitem{copelandetal}
E.J. Copeland, E.W. Kolb, A.R. Liddle and J.E. Lidsey, Phys. Rev. D {\bf 48}, 2529 (1993).

\bibitem{dingle}
H. Dingle, Mon. Not. Roy. Astro. Soc. {\bf 94}, 134 (1933).

\bibitem{eddington1}
A.S. Eddington, {\it The Expanding Universe}, Cambridge (1933).

\bibitem{eddington2}
A.S. Eddington,  Mon. Not. Roy. Astro. Soc. {\bf 90}, 668 (1930).

\bibitem{ehlersbuchert}
J. Ehlers and T. Buchert, Gen. Rel. Grav. {\bf 29}, 733 (1997).

\bibitem{ehlersrindler}
J. Ehlers and W. Rindler, Mon. Not. Roy. Astro. Soc. {\bf 238}, 503 (1989).

\bibitem{einstein}
A. Einstein, {\it Sitzungsbericht Preuss. Akad. Wiss. Berlin}, pp.142--152 (in German) 
(1917).

\bibitem{ellis}
G.F.R. Ellis,  in {\em General Relativity and Gravitation} (D. Reidel
Publishing Co., Dordrecht), pp.\ 215--288 (1984).

\bibitem{ellis:dahlem}
G.F.R. Ellis, G. B\"orner, T. Buchert, J. Ehlers, C.J. Hogan, R.P. Kirshner, W.H. Press, 
G. Raffelt, F--K. Thielemann and S. Van den Bergh, 
 in: {\em Dahlem Workshop Report ES19 The Evolution of the Universe}, 
Berlin 1995, G. B\"orner and S. Gottl\"ober (eds.), Chichester: Wiley, pp.51--78 (1997).

\bibitem{ellisbruni}
G.F.R. Ellis and M. Bruni, Phys. Rev. D {\bf 40}, 1804 (1989).

\bibitem{ellisbuchert}
G.F.R. Ellis and T. Buchert, Phys. Lett. A. (Einstein Special Issue) {\bf 347}, 38 (2005).

\bibitem{ellis:emergent1}
G.F.R. Ellis and R. Maartens, Class. Quant. Grav. {\bf 21}, 223 (2004).

\bibitem{ellis:emergent2}
G.F.R. Ellis, J. Murugan and C.G. Tsagas, Class. Quant. Grav. {\bf 21}, 233 (2004).

\bibitem{ellisstoeger}
G.F.R. Ellis and W. Stoeger, Class. Quant. Grav. {\bf 4}, 1697 (1987).

\bibitem{evans}
A.K.D. Evans, I.K. Wehus, {\O}. Gr{\o}n and {\O}. Elgan{\o}y, 
Astron. Astrophys. {\bf 430}, 399 (2005).

\bibitem{futamase1}
T. Futamase, Mon. Not. Roy. Astro. Soc. {\bf 237}, 187 (1989).

\bibitem{futamase2}
T. Futamase, Phys. Rev. D {\bf 53}, 681 (1996).

\bibitem{brandenberger2}
G. Geshnizjani and R.H. Brandenberger, Phys. Rev. D {\bf 66}, 123507 (2002).

\bibitem{brandenberger3}
G. Geshnizjani and R.H. Brandenberger, J.C.A.P. {\bf 04} , 006 (2005).

\bibitem{gibbons87}
G.W. Gibbons, Nucl. Phys. B {\bf 292}, 784 (1987).

\bibitem{gibbons88}
G.W. Gibbons, Nucl. Phys. B {\bf 310}, 636 (1988).

\bibitem{giovanellietal}
R. Giovanelli, D.A. Dale, M.P. Haynes, E. Hardy and L.E. Campusano, 
The Astrophys. J. {\bf 525}, 25 (1999).

\bibitem{harrison}
E.R. Harrison, Rev. Mod. Phys. {\bf 39}, 862 (1967).

\bibitem{harrison:book}
E.R. Harrison, 
{\it Cosmology: The Science of the Universe}, Cambridge Univ. Press (1981; 2000). 

\bibitem{hellaby}
C. Hellaby, Gen. Rel. Grav. {\bf 20}, 1203 (1988).

\bibitem{hosoya:infoentropy}
A. Hosoya, T. Buchert and M. Morita, Phys. Rev. Lett. {\bf 92}, 141302 (2004).

\bibitem{hubble}
E. Hubble, Ap. J. {\bf 79}, 8 (1934).

\bibitem{wald}
A. Ishibashi and R.M. Wald, Class. Quant. Grav., in press (2005).

\bibitem{kasai}
M. Kasai, Phys. Rev. D {\bf 52}, 5605 (1995).

\bibitem{kolbetal}
E.W. Kolb, S. Matarrese and A. Riotto, submitted to Phys. Rev. D; 
astro--ph/0506534 (2005). 

\bibitem{langloisvernizzi}
D. Langlois and F. Vernizzi, Phys. Rev. D {\bf 72}, 103501 (2005).

\bibitem{lehoucq:topology}
R. Lehoucq, J.--P. Uzan and J.--P. Luminet, Astron. Astrophys. {\bf 363}, 1 (2000). 

\bibitem{lemaitre33a}
A. Lema\^\i tre,  {\it CR} {\bf 196}, p.903 and p.1085 (1933a). 

\bibitem{lemaitre33b}
A. Lema\^\i tre, Ann. Soc. Sci. Bruxelles {\bf A53}, 51 (1933b).

\bibitem{linde1}
A.D. Linde, D.A. Linde and A. Mezhlumian, Phys. Rev. D {\bf 49}, 1783 (1994). 

\bibitem{linde2}
A.D. Linde, D.A. Linde and A. Mezhlumian, Phys. Rev. D {\bf 54}, 2504 (1996).

\bibitem{unruh:static}
B. Losic and W.G. Unruh, Phys. Rev. D {\bf 71}, 044011 (2005).

\bibitem{maartensetal1}
R. Maartens, G.F.R. Ellis and W.R. Stoeger, Phys. Rev. D. {\bf 51}, 1525 and 5942 (1995).

\bibitem{maartensetal2}
R. Maartens, G.F.R. Ellis and W.R. Stoeger, Astron. Astrophys. {\bf 309}, L7 (1996). 

\bibitem{lahav:virial}
I. Maor and O. Lahav, J. C. A. P. {\bf 07}, 003 (2005).

\bibitem{moffat}
J.W. Moffat, astro--ph/0505326 v4 (2005).

\bibitem{nambu}
Y. Nambu and M. Tanimoto, gr--qc/0507057 (2005).

\bibitem{peebles}
P.J.E. Peebles, {\it The Large Scale Structure of the Universe}, 
Princeton Univ. Press (1980).

\bibitem{peeblesratra}
P.J.E. Peebles and B. Ratra, Rev. Mod. Phys. {\bf 75}, 559 (2003).

\bibitem{piazza}
F. Piazza and S. Tsujikawa, J.C.A.P. {\bf  07}, 004 (2004).

\bibitem{prigogine}
I. Prigogine, {\sl Non Equilibrium Statistical Mechanics}, Interscience Pub., 
New York (1962). 

\bibitem{rasanen:darkenergy}
S. R\"as\"anen,  J.C.A.P. {\bf 02}, 003 (2004).

\bibitem{rasanen:LTB}
S. R\"as\"anen,  J.C.A.P. {\bf 11}, 010 (2004).

\bibitem{rasanen:constraints}
S. R\"as\"anen, submitted to Class. Quant. Grav..; astro--ph/0504005 (2005).

\bibitem{russetal}
H. Russ, M.H. Soffel, M. Kasai and G. B\"orner, Phys. Rev. D {\bf 56}, 2044 (1997).

\bibitem{schmidt04}
B.P. Schmidt, Bull. Astr. Soc. of India {\bf 32}, pp.269--281 (2004).

\bibitem{sota:RG}
Y. Sota, T. Kobayashi, K. Maeda, T. Kurokawa, M. Morikawa and A. Nakamichi,
Phys. Rev. D {\bf 58}, 3502 (1998).

\bibitem{WMAP}
D.N. Spergel et al., The Astrophys. J. Suppl. {\bf 148}, 175 (2003).

\bibitem{tanimoto}
M. Tanimoto, Prog. Theor. Phys. {\bf 102}, 1001 (1999).

\bibitem{tolman}
R.C. Tolman, Proc. N.A.S. {\bf 20}, 169 (1934).

\bibitem{tolman39}
R.C. Tolman, Phys. Rev. {\bf 55}, 364 (1939).

\bibitem{tomita1}
K. Tomita, The Astrophys. J. {\bf 529}, 26 (2000).

\bibitem{tomita2}
K. Tomita, The Astrophys. J.  {\bf 529}, 38 (2000).

\bibitem{tomita3}
K. Tomita, Mon. Not. Roy. Astro. Soc. {\bf 326}, 287 (2001).

\bibitem{uzan:topology}
J.--P. Uzan, R. Lehoucq and J.--P. Luminet, Astron.\ Astrophys. {\bf 351}, 766 (1999). 

\bibitem{wetterich}
C. Wetterich, Phys. Rev. D {\bf 67}, 043513 (2003).

\bibitem{wiltshire}
D.L. Wiltshire, gr--qc/0503099 (2005).

\bibitem{zal97}
R.M. Zalaletdinov,
Bull. Astron. Soc. India {\bf 25}, 401 (1997).

\bibitem{zehavietal}
I. Zehavi, A.G. Riess, R.P. Kirshner and A. Dekel, 
The Astrophys. J. {\bf 503}, 483 (1998).

\end{thebibliography}
\end{document}